\documentclass[journal]{IEEEtran}

\usepackage{setspace}
\usepackage{lettrine}

\usepackage{graphics,
	psfrag,
	epsfig,
	amsthm,
	cite,
	amssymb,
	url,
	dsfont,
	subfigure,
	algorithmic,
	balance,
	enumerate,
	color,
	setspace
}
\usepackage{amsmath}

\usepackage{epstopdf}

\newcommand{\eref}[1]{(\ref{#1})}
\newcommand{\sref}[1]{Section~\ref{#1}}

\newcommand{\cref}[1]{Constraint~\ref{#1}}

\newcommand{\ignore}[1]{}

\IEEEoverridecommandlockouts
\usepackage{mathtools}
\usepackage{color}
\usepackage[utf8]{inputenc}
\usepackage[ruled, lined, longend]{algorithm2e}
\SetEndCharOfAlgoLine{}

\begin{document}
	
\IEEEoverridecommandlockouts


\title{\vspace{-.9cm} Task Offloading Optimization in  NOMA-Enabled Multi-hop Mobile Edge Computing System Using  Conflict Graph}

\title{\vspace{-.9cm} Task Offloading Optimization in  NOMA-Enabled Multi-hop Mobile Edge Computing System Using  Conflict Graph}

\author{
	\IEEEauthorblockN{Mohammed S. Al-Abiad, \textit{Student Member, IEEE}, Md. Zoheb Hassan, \textit{Student Member, IEEE}, and Md. Jahangir Hossain, \textit{Senior Member, IEEE}}
	
	\thanks {
		Mohammed S. Al-Abiad and Md. Jahangir Hossain are with the School of
		Engineering, University of British Columbia, Kelowna, BC V1V 1V7, Canada
		(e-mail: m.saif@alumni.ubc.ca, jahangir.hossain@ubc.ca).
		
		Md. Zoheb Hassan is with $\acute{\text{E}}$cole de technologie sup$\acute{\text{e}}$rieure (ETS), University of Quebec, Canada (e-mail:  md-zoheb.hassan.1@ens.etsmtl.ca).
	}
	\vspace{-0.4cm}
}

\maketitle
\begin{abstract}
	
	Resource allocation is investigated for offloading computational-intensive tasks in  multi-hop mobile edge computing (MEC) system. The envisioned system has both the cooperative access points (AP) with the computing capability and the MEC servers. A user-device (UD) therefore first uploads a computing task to the nearest AP, and the AP can either locally process the received task or offload to MEC server. In order to utilize the radio resource blocks (RRBs) in the APs efficiently, we exploit the non-orthogonal multiple access for offloading the tasks from the UDs to the AP(s). For the considered NOMA-enabled multi-hop MEC computing system, our objective is to minimize both the latency and energy consumption of the system jointly. Towards this goal, a joint optimization problem is formulated by taking the offloading decision of the APs, the scheduling among the UDs, RRBs, and APs, and UDs' transmit power allocation into account. To solve this problem efficiently,   (i) a conflict graph-based approach is devised that solves the scheduling among the UDs, APs, and RRBs, the transmit power control, and  the APs'  computation resource allocation  jointly, and (ii) a low-complexity   pruning graph-based approach is also  devised.  The efficiency of the  proposed graph-based approaches over several benchmark schemes is verified via extensive simulations.

\end{abstract}


\section{Introduction} \label{sec:I}
The emerging Fifth-Generation (5G) era has brought tremendously increasing computational-intensive and energy-demanding applications, such as 3D modeling and online gaming. However, high computation capability and processing power are required to execute such advanced applications. Today's smart mobile devices  are still resource-constrained in terms of battery and computational capacity. Accordingly, these resource constraints present a key barrier to execute the computational-intensive applications at the mobile devices \cite{1}. Recently,  Mobile Edge Computing (MEC) has emerged as an enabling technology to improve computation performance of the 5G era \cite{2}. In an MEC enabled system, powerful MEC servers are placed in close proximity to UDs \cite{4,5}.  Specifically, MEC enables executing the computational-intensive tasks  at the network edge servers, thanks to the increased intelligence of the network edge and improved communication performance between mobile devices and network edge servers. This reduces the computation burden and energy consumption of the mobile devices \cite{2,3}.

With the recent progress in fog radio access network, access points (APs) also have certain computation capability \cite{FRAN}. Therefore, MEC can exploit computational capabilities of  APs in the network edge and MEC servers. This makes MEC a suitable solution for offloading the computational-intensive and time-sensitive mobile applications. On the one hand, by offloading the computing tasks to adjacent MEC servers \cite{6,7}, the low-latency and high-reliable applications can be executed at the UDs efficiently. On the other hand,  the energy consumption of the mobile devices can be improved \cite{8}, and consequently, the battery life of smart devices can be prolonged \cite{9}. However, to fully capitalize the advantage of the MEC system, the performance of task offloading from the mobile UDs to MEC servers needs to be optimized. Specifically, the performance of task offloading strongly depends on the data transmission capacity of the links between UDs and MEC servers. Hence, smart resource allocation strategy is imperative for the MEC enabled system, especially  in dense networks with large number of UDs and limited radio resource blocks (RRBs) \cite{10}.

Non-orthogonal multiple access (NOMA) \cite{11,12} can  improve the spectrum efficiency of a cellular network by increasing the number of scheduled  UDs per RRB. Therefore, NOMA is a potential solution to address the challenging problem of supporting the increased number of UDs with the limited number of RRBs in beyond-5G era systems \cite{13}. Specifically, in uplink NOMA, multiple UDs are multiplexed over the same RRB, and the access point (AP) sequentially decodes the UDs' messages by implementing a successive interference cancellation (SIC) technique \cite{14,15}. Note that the task offloading from the UDs to the MEC server is an uplink multiple access scheme. Consequently, leveraging the improved spectral efficiency and system capacity, NOMA is also a potential solution to reduce the latency and enhance the energy efficiency of the task offloading process in an MEC system \cite{14,15}.  Accordingly, developing an innovative resource allocation framework is imperative for harnessing the aforementioned benefits of NOMA-based MEC  systems.

\subsection{Related Works and Motivation}
Motivated by their numerous potential advantages, MEC and NOMA have attracted extensive studies in their own area \cite{AA}, \cite{12} and references therein, respectively. Recently, NOMA-based MEC has achieved considerable attention to optimize two main metrics, namely, (i) energy consumption minimization \cite{16, 17, 18, 19, 20, 21, 22, 23} and  (ii) delay minimization \cite{10, 14, 24, 25}. The existing works that are separately and jointly optimized these two metrics  are explained as follows.

\textit{Related works for energy consumption minimization:} In \cite{16}, \cite{19}, the authors optimized the transmit time and power consumption in NOMA-based MEC uplink and downlink systems using geometric programming and successive convex approximation, respectively. In \cite{20}, the authors jointly optimized the offloading decision, radio resource allocation, and the SIC decoding to minimize the latency weighted system energy minimization in NOMA-based MEC systems. In \cite{21},
the authors jointly considered radio and computation resource allocation and  developed a partial task offloading scheme in a NOMA-based MEC heterogeneous network. Particularly, the authors in \cite{21} minimized the energy consumption given a maximum tolerable latency. In \cite{22},  the authors  investigated joint communication and computation resource allocation in a NOMA-based MEC system while assuming all the UDs offload their tasks to the MEC server. In \cite{23}, the authors developed a genetic algorithm based strategy for minimizing the total transmit power in a wireless network.

\textit{Related works for delay minimization:} Besides minimizing energy consumption, it is also imperative to reduce latency  for the delay-sensitive applications. To this end, the following studies investigated resource allocation to minimize latency in NOMA-based MEC system.  The authors of \cite{14} minimized the delay of task offloading in a NOMA-enabled
MEC system by first transforming the delay-minimization problem into a fractional programming problem, and then developed two iterative solutions. Different form \cite{14}, the authors of \cite{24} minimized the maximum task execution latency in a MEC-based NOMA system by optimizing jointly the  SIC ordering and computation resource allocation. For improving the efficiency of wireless data transmission, and hence, minimizing the task offloading delay, the authors of \cite{25} exploited NOMA-enabled multi-access MEC by jointly considering the UDs’ offloaded workloads and the NOMA transmission duration. The authors of \cite{10}  developed a NOMA-enabled partial computation offloading scheme, where NOMA was employed for both offloading the tasks  and  downloading the computed results to and from the MEC server, respectively. By designing resource allocations for both uplink and downlink jointly, the  authors of \cite{10} minimized the overall delay of completing all the UDs’ tasks.

\textit{Related works for joint energy consumption and delay minimization:} In the recent literature, a joint optimization of energy and delay metrics was also studied. 
The authors of \cite{26} minimized the weighted sum of the total delay and energy consumption of all the UDs  in a NOMA-enabled fog-cloud computing
system.  For the same objective, the authors of \cite{28} jointly optimized the computation resource, power and subcarrier allocations, and an iterative algorithm was developed. The authors of \cite{27} systematically investigated the benefits of using NOMA in MEC for both uplink and downlink transmissions and demonstrated that the task processing latency and energy consumption can be minimized by introducing NOMA into MEC systems. Although the aforementioned studies exploited the integration of MEC and NOMA extensively to optimize both latency and energy consumption, the multi-hop MEC architecture with both intelligent APs and MEC servers was not considered in the literature.  We emphasize that such a multi-hop MEC architecture is a promising framework to extend the computation service for the cell-edge UDs while exploiting the computation capability of both APs and MEC servers efficiently.  Moreover, given the limited number of RRBs compared to number of UDs, NOMA is also promising to improve the number of supported UDs in such a multi-hop MEC architecture. Consequently,  in this paper, we study a novel system that integrates the advantages of both multi-hop MEC and NOMA approaches for providing computation services to a large number of UDs.

\textit{Motivation:} The aforementioned works primarily considered a single-hop scenario where the UDs' tasks can only be processed at the MEC servers. However, such a single-hop scenario is not efficient for offloading tasks from the cell-edge UDs. In other words, multi-hop scenario needs to be considered for providing reliable computation services to the cell-edge UDs. Motivated by such a requirement,  in this work, we investigate NOMA-enabled multi-hop MEC system where the UDs' tasks can be processed locally at APs or offloaded to MEC servers. Particularly, in the considered system,  a set of UDs  transmit computational-intensive tasks to a given set of APs where  each AP is equipped with number of RRBs. We emphasize that the notion of APs with computation capability is consistent with the recent progress in fog radio access network.  Moreover, the architecture of cooperative APs with many RRBs was adopted for  maximizing the cloud offloading \cite{S1}, maximizing the sum-rate \cite{S2}, and minimizing the delay \cite{S3} in both fog and cloud radio access networks.  Inspired by \cite{S1, S2, S3}, this work develops a framework to increase the number of collected tasks from the UDs either for local processing at APs or for offloading them to MEC servers while minimizing  both the task processing delay and the energy consumption simultaneously.

\textit{Challenges:} This paper attempts to solve a joint latency-energy minimization problem in a NOMA-enabled multi-hop MEC system. By applying uplink NOMA protocol over the RRBs, the system can  remarkably increase the number of admitted UDs to the  APs, and thus, the proposed system is suitable for providing the computation service in a dense network.
However, the following  challenges need to be addressed. First of all, to take advantage of a NOMA setting for task offloading, the association among the UDs, RRBs, and APs needs to be determined efficiently. Moreover,   NOMA inevitably introduces interference in the system, and thus, the transmit power of the UDs needs to be optimized to further improve the transmission rate and reduce latency of task offloading process. Finally,  the APs usually do not have enough computing resources and may not meet large-scale UDs' tasks. Thus, the APs need to smartly offload large-scale UDs' tasks to adjacent MEC servers. Nevertheless, the multi-hop communication can introduce additional latency.  Essentially, the computation resources of both the MEC servers and the APs need to be managed appropriately to take advantage of both the local and centralized computation services. It is therefore evident that  the NOMA-enabled  multi-hop MEC system has several degrees-of-freedom, namely,  the UD-AP-RRB association, UD power optimization, local computations at the APs, and task offloading at the MEC servers.  A joint optimization of the aforementioned degrees-of-freedom is presented. To the best of the authors' knowledge, this is the first attempt in the literature to optimize jointly the degrees-of-freedom of a NOMA-enabled multi-hop MEC system.

\subsection{ Contributions}
In this paper, we consider incorporating multi-hop task processing into the conventional MEC-NOMA system.  In the envisioned system, the APs have multiple RRBs that collect tasks for local processing at the APs or for offloading to MEC servers. To this end, we introduce innovative graph-theoretical frameworks by taking  the UD-AP-RRB scheduling, UD power optimization, local computation resource allocation, admission control and offloading decisions into account. 
The main contributions of our work are presented as follows. 
\begin{enumerate}
	\item For a NOMA-enabled multi-hop MEC system, we develop a framework where APs and MEC servers collaborate to minimize the delay of processing tasks and energy consumption jointly. To this end, a weighted-sum method of latency-energy consumption minimization problem is formulated with the constraints on UD-AP-RRB scheduling, maximum local computation resource allocation, UD transmission power, and maximum tolerable delay guarantee. Such an optimization problem is NP-hard
	and computationally intractable. To obtain a tractable solution, we decompose the original problem into two sub-problems, and obtain efficient solutions to both sub-problems. The first subproblem  obtains  scheduling among UDs, RRBs, and APs, UDs' transmit power control, and local computation resource allocation. The second sub-problem obtains the decisions of offloading the collected tasks from the APs to MEC server(s). 
	
   \item To solve the first sub-problem, we design a joint MEC (J-MEC) graph that optimally solves the UD-AP-RRB scheduling and power allocation problem, jointly, and we propose a closed-form solution for the local computation resource allocation for the given UD-AP-RRB scheduling. To solve the second sub-problem, we propose a cross-layer weighting solution based on the optimized resource allocation variables of the first sub-problem. The aforementioned graph-based and cross-layer weighting solutions of the corresponding sub-problems is referred to a \textit{Joint Approach}.
   
	\item Since the joint approach requires generating all the possible NOMA clusters (i.e., combinations between UDs), it exhibits high computational complexity. Therefore, we develop a low-complexity graph pruning algorithm that judiciously generates NOMA clusters that are feasible for tasks local processing. Using these generated NOMA feasible clusters, our proposed algorithm designs a new reduced NOMA graph. Thereafter, by applying a greedy maximum-weight-independent set (MWIS) algorithm, we obtain a feasible solution to the reduced NOMA graph.

	\item Numerical  results reveal that the proposed schemes offer improved latency and energy consumption as	compared to the baseline schemes. 	Moreover, our proposed pruning graph approach reduces the computational complexity significantly with small performance loss compared to the joint approach.
	

\end{enumerate}

The rest of this paper is organized as follows. The system model is described  in  \sref{SMMM}.  We formulae and transform the weighted-sum latency-energy consumption optimization problem in \sref{PF}. In \sref{G} and \sref{MT}, we develop a joint approach and a low-complexity graph pruning solution to find the optimized scheduling decision, respectively. Simulation results are presented in
\sref{NR}, and in \sref{C}, we conclude the paper.

\section{System Model} \label{SMMM}
\ignore{In this section, we will start with a detailed description of our system model, and then provide the local processing and task offloading.}

\begin{figure}[t!]
	\centering
	\includegraphics[width=0.75\linewidth]{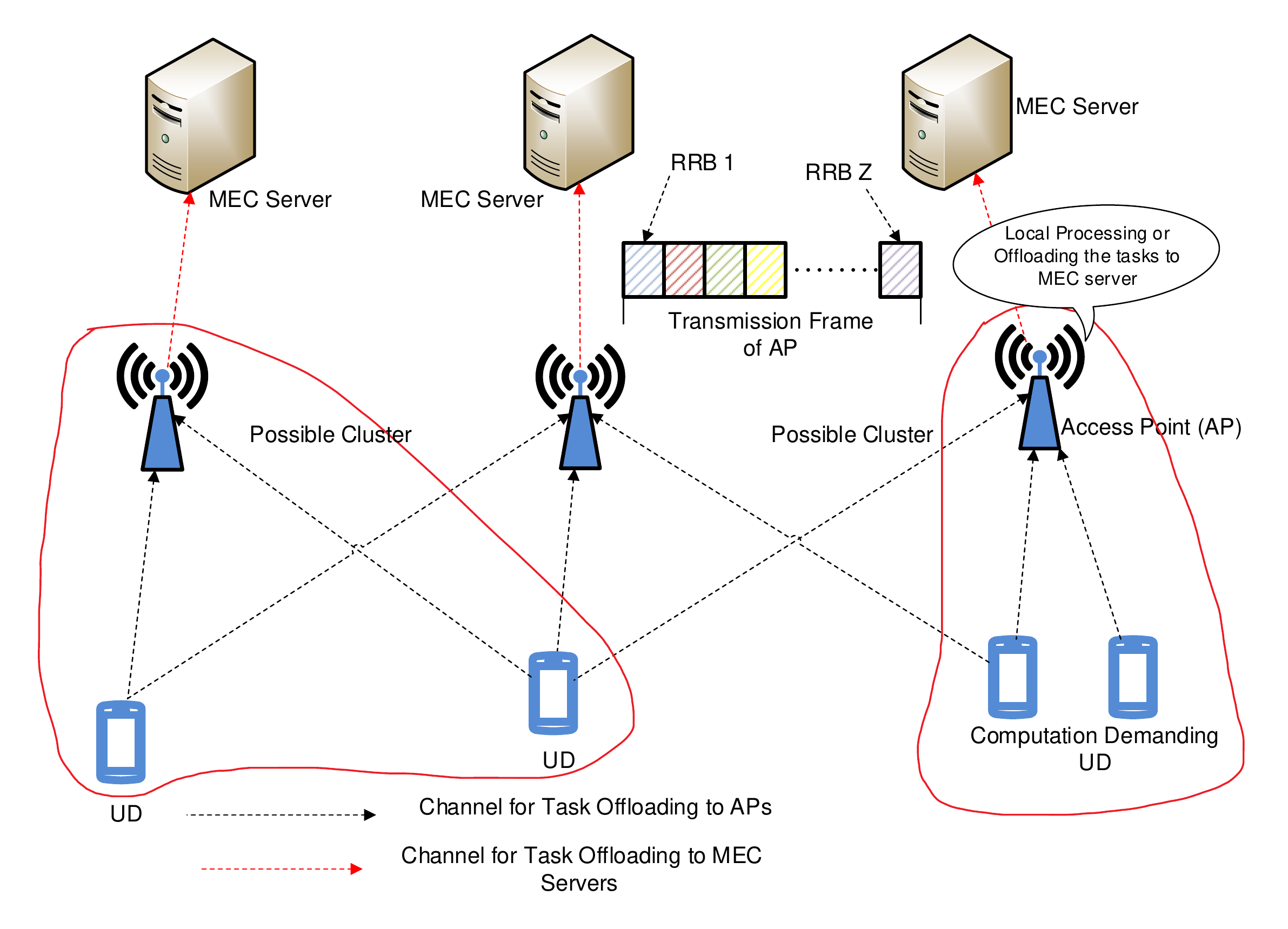}
	\caption{Multi-hop MEC resource setting with $3$ Access Points (APs).}
	\label{fig1}
\end{figure}
\subsection{System Overview} 
We consider the NOMA-enabled and multi-hop MEC system  illustrated in Fig. \ref{fig1} with $K$ MEC servers, $M$ access points (APs), and $N$ computation demanding user devices (UDs). The sets of MEC servers, APs, and UDs are denoted
by  $\mathcal{K}=\{1,2,\cdots,K\}$,  $\mathcal{M}=\{1,2,\cdots,m\}$,  and  $\mathcal{N}=\{1,2,\cdots,n\}$, respectively. MEC servers and APs are equipped with CPUs that help computing the tasks, and accordingly, they cooperate with each other to execute the demanding computationally-intensive tasks of $N$ UDs. The MEC servers possess
stronger task processing capabilities, while the APs
have certain computation resources \cite{FRAN}, \cite{AP}. 
We consider that each AP has a limited coverage range, denoted by $\mathcal S_{m}$, which represents the service area of the $m$-th AP within a circle of radius $\mathtt R$. The service area of each AP is a set of UDs which defined by $\mathcal S_{m}=\{n\in \mathcal{N}| d_{m,n}\leq \mathtt R$\}, where $d_{m,n}$ is the distance between the $m$-th AP and the $n$-th UD. Let $\mathbf C=\{c_{m,n}\}$ be the AP allocation matrix, where element $c_{m,n}=1$ represents that the $n$-th UD is allocated to the $m$-th AP, and $c_{m,n}=0$ otherwise.  \ignore{ Accordingly, we group APs and their reachable UDs into clusters. Thus, each AP and its reachable UDs can form a cluster $Z$. Let $\mathbf C$ denote the set of all the clusters in the network which can be expressed as $\mathbf C=\{C_1 ,C_2,\ \cdots, C_m\}$, $C_i \subset \mathcal N, \forall i,j \in\{1,2,\ \cdots,m\}$.}  

Similar to \cite{S1, S2}, we consider that each AP has $Z$ orthogonal radio resource block (RRBs) that are denoted by the set $\mathcal{Z}=\{1,2,\cdots,Z\}$, where UDs can transmit their demanding tasks to the APs for local processing or for offloading to MEC servers via APs. Thus, the total number of RRBs in the system is $Z_\text{tot}=ZM$. Let $\mathbf R=\{r^n_{z,m}\}$ be the RRB allocation matrix, where element $r^n_{z,m}=1$ represents that the $n$-th UD is allocated to the $z$-th RRB in the $m$-th AP, and $r^n_{z,m}=0$ otherwise. 
We assume each UD can be allocated with only one RRB, and to schedule number of UDs to each RRB, we consider NOMA. To manage co-channel interference caused by NOMA, the APs utilize SIC technique to decode the multiple overlapping signals sequentially, where the decoding order is descending in terms of channel power gain. For reducing the complexity of SIC-based decoding, we consider that each RRB can only serve at most $2$ UDs. Consequently, the number of admitted UDs for task processing is at most $2Z_\text{tot}$, and the rest local-infeasible and not-admitted UDs will fail in task processing. We denote that the 2-UDs association in each RRB by a NOMA cluster.

On the UD side, we consider that each UD $n$ has only one inseparable computation
task, which can be represented by $t_n=\{B_n, \lambda_n, T_{\max}^n\},~\forall n \in \mathcal N$, where $B_n$ is the size of input data (in bits), $\lambda_n$ is the processing density (in CPU cycles/bit), and $T_{\max}^n$ is the maximum tolerable
latency (in second) within which the task should be completed
successfully. The input data of the $n$-th UD with a size of $B_n$ should be locally processed at the potential AP or transferred to the MEC server in task offloading mode \cite{33, 34}. Since tasks have different sizes, they need an efficient computational processing strategy such that they can be processed within their maximum tolerable latency.

For UD-AP uplink transmission strategy, tasks are transmitted
over different RRBs via a wireless link. Let $P_n$ and $Q_{m}$ denote the transmission power of the $n$-th UD and the $m$-th AP, respectively. Let $\bf P$ be a $1\times N$ matrix
containing the power levels of all UDs, i.e., 
$\textbf P = [P_n]$. To avoid complexity, we consider fixed transmission power for the APs. The instantaneous signal-to-interference-plus-noise (SINR)  for the link between the $n$-th UD and the $z$-th RRB in the $m$-the AP is given by
\begin{align} 
\gamma^n_{m,z}=\frac{r^n_{z,m}P_n \left|h^n_{m,z}\right|^2}{\sum_{\substack{n' \in \mathcal{N}, (n',n) \in \mathcal A_m\\h^{n'}_{m,z}<h^n_{m,z}}}r^{n'}_{z,m}P_{n'} \left|h^{n'}_{m,z}\right|^2 +\sigma^2},
\end{align} 
where $\sigma^{2}$ denotes the additive white Gaussian noise variance and $h^n_{m,z}$ denotes the channel fading gain for the link between the $n$-th UD and the $z$-th RRB in the $m$-th AP. Then, the transmit rate of the $n$-th UD on the $z$-th RRB in the $m$-th AP can be given by $R^n_{m,z}=B_0\log_{2}(1+\gamma^n_{m,z})$, where $B_0$ is the bandwidth of the $z$-th RRB. Consequently, the transmit rate of UD $n$ can
be given by
\begin{align} 
R_n=\sum_{m\in \mathcal M}\sum_{z\in \mathcal Z}B_0\log_{2}(1+\gamma^n_{m,z}).
\end{align} 

At the APs, UDs' collected tasks can be either processed locally or offloaded to MEC servers. Therefore, APs offload UDs' tasks (if not processed locally) to MEC servers via multiple orthogonal channels. So, the transmission rate of the $m$-th AP to offload the collected data to the $k$-th MEC server is
$C_{k}^{\text{m}}= \log_{2}\left(1+\cfrac{Q_m|h_{m,k}|^{2}}{\sigma^2}\right),$ where $h_{m,k}$ denotes the channel gain from the $m$-th AP to the $k$-th MEC server.

\subsection{Local Processing and Task Offloading}
\textit{1) Local processing:} Let $f^\text{loc}_m$ be the computational
speed of the CPU in the $m$-th AP (in cycles per second). Let $\bf f^\text{loc}$ be a $1\times M$ matrix
containing the local computations of all APs, i.e., 
$\textbf f^\text{loc} = [f^\text{loc}_m]$. The uplink transmission delay for sending the $t_n$-th task to the $z$-th RRB in  the $m$-th AP is $U^n_{z,m}=\frac{B_n}{R^{n}_{m,z}}$. Let $\tau_m=\{t_1, t_2, \cdots, t_{|\tau_m|}\}$ denote the set of collected tasks at the $m$-th AP across all the RRBs. Then, the delay (i.e., uploading transmission delay and task processing delay) and energy consumption of $\tau_m$ in local processing mode can be given by $T^\text{loc}_{\tau_m}=\max_{n \in \tau_m} \{U^n_{z,m}\}+\frac{( \sum^{|\tau_m|}_{n=1}B_n)\lambda_n}{f^\text{loc}_m}$ and $E^\text{loc}_{\tau_m}=\alpha ( \sum^{|\tau_m|}_{n=1}B_n)\lambda_n (f^\text{loc}_m)^2$, respectively, where\ignore{where $U_{z,m}=\max(\frac{B_n}{R^{n}_{m,z}}, \frac{B_{n'}}{R^{n'}_{m,z}})$}  $\alpha$ is a constant coefficient about the CPU chip architecture \cite{35}. Simply, if AP $m$ has only one RRB and can only collect one task, the delay and energy consumption of task $t_n$ of  UD $n$ in local processing mode can be simply expressed by $T^\text{loc}_{t_n}=U^n_{z,m}+\frac{B_n\lambda_n}{f^\text{loc}_m}$ and $E^\text{loc}_{t_n}=\alpha B_n \lambda_n (f^\text{loc}_m)^2$, respectively.

\textit{2) Task offloading:} If tasks $\{t_n\}_{n\in \mathcal N}$ are going to be processed
at the MEC servers, APs need to transfer the collected
data to the MEC servers. Let
$x_n$ denote the offloading decision of each AP $m$, where
$x_m=1$ indicates that the collected tasks $\tau_m$ is offloaded, and $x_m=0$
otherwise. We define the offloading decision
$\mathbf X=[x_{m,k}]_{\mathcal M \times \mathcal K}$ such that:
\begin{equation}
x_{m,k} = \begin{cases}
1 &\text{if collected tasks $\tau _m$ are offloaded to MEC server $k$,}\\
0 & \text{otherwise.}
\end{cases}
\end{equation}
Based on the offloading decision $\mathbf X$, the uplink transmission delay for sending the collected tasks $\tau_m$ from the $m$-th AP to the $k$-th MEC server is $U^\text{MEC}_{m,k}=\sum^K_{k=1} x_{m,k}\frac{\sum^{|\tau_m|}_{n=1}B_n}{R^{m}_{k}}$. \ignore{ via orthogonal subchannels for parallel offloading scheme and via a shared channel for sequential offloading scheme.}
Let $f^\text{MEC}_k$ be the computational
speed of the CPU in the $k$-th MEC server (in cycles per second). For analytical tractability, we consider that MEC server will start
to process $\tau_m$ only when it has successfully received all the corresponding data from the $m$-th AP.  Suppose that MEC server $k$ can allocate the collected tasks $\tau_m$ with sufficient computation
resource $f^\text{MEC}_k$ for task processing, then the delay and energy consumption of tasks $\tau_m$ in task offloading model can be given by
\begin{equation}\label{eq44}
T^\text{MEC}_{\tau_m}=\max_{n \in \tau_m} \{U^n_{z,m}\}+\frac{\sum^{|\tau_m|}_{n=1}B_n}{R^m_k}+\frac{( \sum^{|\tau_m|}_{n=1}B_n)\lambda_n}{f^\text{MEC}_k},
\end{equation}

\begin{equation}\label{eq55}
E^\text{MEC}_{\tau_m}= \frac{\sum^{|\tau_m|}_{n=1}B_n}{R^m_k}Q_m+\frac{( \sum^{|\tau_m|}_{n=1}B_n)\lambda_n}{f^\text{MEC}_k}Q^{idle}_{m},
\end{equation}
where $Q^{idle}_{m}$ is the idle power of the $m$-th AP. In \eref{eq44}, the first term represents the transmission duration for sending $\tau_m$ from UDs to the $m$-th AP, the second term represents the transmission duration for sending $\tau_m$ from the $m$-th AP to the $k$-th MEC server, and the third term is for task processing at the $k$-th MEC server.

\textit{3) Offloading scheme:} The collected tasks of the UDs can be computed locally at the APs or simultaneously offloaded
to the MEC servers from the APs  via orthogonal channels. The scheduling problem of this offloading scheme
consists of assigning tasks to APs or MEC servers under the following constraints:
\begin{itemize}
	\item Each UD is assigned to only one RRB and one AP, while each RRB can schedule at most two UDs using NOMA.
	\item The collected tasks $\tau_m$ can be locally processed at one AP or offloaded to only one MEC server based on the offloading decision $\mathbf X$.
	
\end{itemize}

\textit{4) Admission control and energy consumption:} Since the number of MEC servers is less than that of APs, at most $K$ APs can be admitted in for tasks
offloading. Let $y_m$ be the
admission control variable, where $y_m=1$ denotes that the collected tasks $\tau_m$ at the $m$-th AP is permitted to access the MEC servers, and $y_m=0$
otherwise. By considering offloading decision and
admission control, the delay and energy consumption of
tasks $\tau_m$ is given by
\begin{equation}
T_{\tau_m}=(1-x_{m})T^\text{loc}_{\tau_m}+x_{m}y_mT^\text{MEC}_{\tau_m},
\end{equation}
\begin{equation}
E_{\tau_m}=(1-x_{m})E^\text{loc}_{\tau_m}+x_{m}y_mE^\text{MEC}_{\tau_m}.
\end{equation}
Consequently, the latency of
processing tasks $\tau_m$ can be expressed as 
\begin{equation}
\mathcal D_{\tau_m}(\mathbf X)=T_{\tau_m}(\mathbf X),
\end{equation}
where $T_{\tau_m}(\mathbf X)$ is the computational processing delay of tasks $\tau_m$ given the offloading schedule $\mathbf X$, and accordingly, the total latency is given by 
\begin{equation}\label{latency}
\mathcal L(\mathbf X)=\max_{m\in \mathcal M}\{\mathcal D_{\tau_m}(\mathbf X)\}\ignore{\footnote{In the case of locally proccessing one task at any AP or offlloding it to any MEC server, the total latency of completing all the tasks is given by $\mathcal L(\mathbf X)=\max_{n\in \mathcal N}\{\mathcal D_{t_n}(\mathbf X)\}$, where $\mathcal D_{t_n}(\mathbf X)=T_{t_n}(\mathbf X)$.}}.
\end{equation}
Finally, the overall energy consumption for locally processing the tasks at the APs or offloading them to the MEC servers based on the offloading decision $\mathbf X$ is expressed as $\mathcal E(\mathbf X)=\sum_{m\in \mathcal M}E_{\tau_m}$. The main symbols used throughout this paper are listed in Table \ref{table_1}.

\setlength{\textfloatsep}{0pt}
\begin{table*}[t!]
	\renewcommand{\arraystretch}{0.9}
	\caption{Main Symbols used in the paper}
	\label{table_1}
	\centering
	\begin{tabular}{p{2.1cm}| p{10.8cm}}
		\hline
		\hline
		
		\textbf{Symbol} & \textbf{Definition}\\
		\hline
		$\mathcal{N}, \mathcal M, \mathcal K, \mathcal Z$ & Sets of $N$ UDs, $M$ APs, $K$ MEC servers, $Z$ RRBs \\
		\hline
		$f^\text{MEC}_k$ & Computational speed of MEC server $k$ (in cycles per second)\\
		\hline
		$f^\text{loc}_m$ & Computational speed of AP $m$ (in cycles per second)\\
		\hline
		$B_{n}$ & Size of task $t_n$ in (bits)\\
		\hline
		$P_n, Q_m$ & Transmission powers of UD $n$ and AP $m$\\
		\hline
		$\mathbf X$ & Offloading decision\\
		\hline
		$\lambda_n$ & Processing density of the task of UD $n$ (in CPU cycles/bit)\\
		\hline
		$T_{\max}^n$ & Maximum tolerable latency of each task (in second) \\
		\hline
		$T^\text{loc}_{\tau_m}$, $E^\text{loc}_{\tau_m}$ & Delay and energy consumption of locally processing tasks $\tau_m$\\
		\hline
		$T^\text{MEC}_{\tau_m}$, $E^\text{MEC}_{\tau_m}$ & Delay and energy consumption of offloading tasks $\tau_m$\\
		\hline
		$\mathcal S_m$ & Set of UDs in the coverage area of AP $m$\\
		\hline
		$\mathbf R$ & Matrix of RRB allocation\\
		\hline
		$R^n_{m,z}$ & Uplink data rate of UD $n$ on RRB $z$ in AP $m$ (in bits/second)\\
		\hline
		$C ^m_{k}$ & Uplink data rate of AP $m$ to MEC server $k$ (in bits/second)\\
		\hline
		$U^n_{z,m}$ & Uplink delay of sending task $t_n$ to RRB $z$ in AP $m$ (in second)\\
		\hline
		$\tau_m$ & Set of collected tasks by AP $m$\\
		\hline 
		$\mathbf C$ & UD-AP allocation matrix\\
		\hline
		$\mathcal L(\mathbf X)$ & Latency of completing all the tasks\\
		\hline
		$\mathcal D_{\tau_m}$& Latency of completing tasks $\tau_m$ (in second)\\
		\hline
		$\mathcal E(\mathbf X)$ & Overall energy consumption\\
		\hline
	\end{tabular}
\end{table*}

\section{Problem Formulation and Transformation}\label{PF}
\subsection{Problem Formulation}
We propose to minimize the completion of processing demanding tasks as well as the energy consumption, and formulate the
problem as the joint optimization of offloading decision $\mathbf X$, RRB
assignment $\mathbf R$, UD-AP association $\mathbf C$, and computational resource allocation. To tackle the trade-off between latency and energy consumption, we consider the weighted sum
method, in which the latency-energy consumption cost function of the scheduling and offloading scheme can be formulated as follows $\pi=\omega_L\mathcal L(\mathbf X)+\omega_E\mathcal E(\mathbf X)$. Here, $\omega_L$ and $\omega_E$ are the predefined weight factors.

Let the binary variable $y_{k,m}$ (where $k \in \mathcal K$ and $m \in \mathcal M$) be $1$ if AP $m$ is scheduled to MEC server $k$, and $0$ otherwise. Consequently, the optimization problem is formulated as $\mathcal{P}_1$ at the top of the next page.
\begin{table*}
	\begin{normalsize} 
\begin{subequations}
	\begin{align} \nonumber \label{eqn:Mschedule1}
& \mathcal{P}_1: 
\min_{\substack{\mathbf X, \mathbf C, \mathbf R, \textbf{f}^{loc},\bf P, \bf y}} \pi\\
&\rm s.t.
	\begin{cases}  \nonumber
\hspace{0.2cm} \text{C1:}\hspace{0.2cm} T_n\leq T_{\max}^n,~\forall n \in \mathcal N, \\
\hspace{0.2cm} \text{C2:}\hspace{0.2cm} \sum_{m\in \mathcal M}c_{m,n} =1 ~\&~ \sum_{z\in \mathcal Z}r^n_{z,m} =1, \forall n \in \mathcal N,\\ 
\hspace{0.2cm} \text{C3:}\hspace{0.2cm} \sum_{n\in \mathcal N}r^n_{z,m} \leq 2, \forall z \in \mathcal Z, m\in \mathcal M,\\
\hspace{0.2cm} \text{C4:}\hspace{0.2cm} \sum_{k\in \mathcal K}y_{k,m} =1, \forall m \in \mathcal M,\\
\hspace{0.2cm} \text{C5:}\hspace{0.2cm} 0 \leq f^{loc}_m \leq f^{loc}_{\max}, ~\forall m\in \mathcal{M},\\
\hspace{0.2cm} \text{C6:}\hspace{0.2cm} 0 \leq P_n \leq P_{\max},\\ \hspace{0.2cm} \text{C7:} \hspace{0.2cm} R_n\geq R_{th}, ~\forall n\in \mathcal{N},\\
\hspace{0.2cm} \text{C8:}\hspace{0.2cm} c_{i,j}\in \{0,1\}, r^k_{i,j}\in \{0,1\}, y_{i,j}\in \{0,1\}, x_{n}\in \{0,1\}.
	\end{cases}
\end{align}
\end{subequations}
	\end{normalsize}
\vspace*{0.1cm}
\hrulefill
\end{table*}
In $\mathcal P_1$, C1 states that each task should be accomplished within a tolerable deadline; C2 indicates that each UD is scheduled to only one AP and allocated with only one RRB; C3 indicates that maximum two UDs can be scheduled to each RRB   at the same time; C4 states that each AP is scheduled to only one MEC server; C5 is the constraint on local computation resource allocation; C6 and C7 are the constraints on transmit power control and rate threshold $R_{th}$, respectively.

\ignore{It is worth noting that the size of the search space for the
optimization problems in $\mathcal P_1$ is $2^{NMKZ}$. Consequently, the
complexity of finding the optimum solution using exhaustive
search is prohibitive for a reasonable number of RDs, MEC servers and tasks. }

\subsection{Problem Transformation}
Solving problem $\mathcal P_1$  owing to its mixed combinatorial characteristics and objective-constraints coupling is challenging \cite{36, 37}.
Hence, it is crucial to propose an effective approach that is affordable for the
large-scale situations of $\mathcal P_1$. To this end, we reformulate $\mathcal P_1$ into two sub-problems, namely, (i) AP side optimization problem that optimizes UD-RRB-AP scheduling, power control, and local computation allocation, and (ii) MEC server side optimization problem that jointly optimizes admission control and offloading decisions. Next, we solve the reformulated sub-problems separately. For the tractability of ensuing the analysis of problem reformulation, we have  the following remarks.

\textit{\textbf{Remark 1:}  In order to maximize the effective system capacity that represents the number of UDs whose tasks are processed successfully, we encourage as many UDs
	to upload their tasks to RRBs and employ NOMA in each RRB.}

\textit{\textbf{Remark 2:} In order to minimize the uplink transmission delay from UDs to APs, we judiciously select a significant set of UDs whose data rates to the RRBs/APs are good, and in each RRB we employ power allocation. }

\textit{Distributed Local Computation Optimization:} Each AP $m$ first assumes that its collected tasks $\tau_m$ are processed locally,
i.e., the offloading decision $x_m=0$, and then solves the problem $\mathcal P_2$, given at the top of the next page, to obtain the UD-RRB-AP scheduling, power optimization, and local resource allocation strategies.
\begin{table*}
	\begin{normalsize} 
\begin{subequations}
	\begin{align}
	& \mathcal{P}_2: 
	\min_{\substack{\textbf{f}^{loc},\bf P, \bf C, \bf R}}\left(\max_{m\in \mathcal M}(T_{\tau_m})+\sum_{m\in \mathcal M}E_{\tau_m}\right)\\
	&\rm s.t.
	\begin{cases}  \nonumber
	\hspace{0.2cm} \text{C2, C3, C5, C6, C7.} \\
	\hspace{0.2cm} \text{C9:}\hspace{0.2cm} \frac{(\sum^{|\tau_m|}_{n=1}B_n)\lambda_n}{f^\text{loc}_m}\leq |\tau_m|T_{\max}^n,~\forall m \in \mathcal M, \\
\end{cases}
	\end{align}
\end{subequations}
\end{normalsize}
\vspace*{0.1cm}
\hrulefill
\end{table*}
In $\mathcal P_2$, the optimization is over the continuous variables $\textbf{f}^{loc}$, $\bf P$, and the discrete variables $c_{m,n}$, and $r^n_{z,m}, \forall m\in \mathcal M, n\in \mathcal N, z\in \mathcal Z$. The multi-variable problem $\mathcal{P}_2$ can be decomposed into the following sub-problems.
\begin{itemize}
	\item \textbf{UD Scheduling and Power Allocation Problem:} For a fixed set of local computation allocation $\textbf{f}^{loc}$,	the optimization problem $\mathcal P_2$ can be written as
	\begin{subequations}
		\begin{align} \nonumber
		& \mathcal{P}_3: 
		\min_{\substack{\bf P, \bf C, \bf R}}\left(\max_{m\in \mathcal M}(T_{\tau_m})+\sum_{m\in \mathcal M}E_{\tau_m}\right)\\
		&\rm s.t. \quad
		\text{C2, C3, C6, C7.}
		\end{align}
	\end{subequations}
	The  problem $\mathcal{P}_3$ is a	 mixed-integer non-linear programming problem, and its suitable solution is obtained by applying the graph theory.
	

	\item \textbf{Local Computation Resource Allocation Problem:} For a fixed UD scheduling and power allocation, the problem $\mathcal P_2$ is independently written per AP as
	\begin{subequations}
		\begin{align} \nonumber
		& \mathcal{P}_4: 
		\min_{\substack{f^{loc}_m}}\left(T_{\tau_m}+E_{\tau_m}\right)\\
		&\rm s.t. \quad 
		\text{C5, C9.}
		\end{align}
	\end{subequations}
Since $\mathcal P_4$ is independent in the objective and constraints, the optimization in each AP is independent from each other, and thus the constraints C1 and C4 in $\mathcal P_4$ can be re-written as $\frac{(\sum^{|\tau_m|}_{n=1}B_n)\lambda_n}{|\tau_m|T_{\max}^n} \leq f^{loc}_m \leq f^{loc}_{\max}, ~\forall m\in \mathcal{M}$. Consequently, $\mathcal{P}_4$ can be farther transferred to the following optimization problem.
	\begin{subequations}
		\begin{align} \nonumber
		& \mathcal{P}_5: 
		\min_{\substack{f^{loc}_m}}\left(T_{\tau_m}+E_{\tau_m}\right)\\
		&\rm s.t.
		\hspace{0.2cm} \hspace{0.2cm} \frac{(\sum^{|\tau_m|}_{n=1}B_n)\lambda_n}{|\tau_m|T_{\max}^n} \leq f^{loc}_m \leq f^{loc}_{\max}, ~\forall m\in \mathcal{M}.
		\end{align}
	\end{subequations}

	\textit{\textbf{Remark 1}: Our local-related optimization has low computational complexity, since the solution to local computation resource allocation $f^{loc}_m$
	can be	obtained in closed form as given in the next section.}
\end{itemize} 

\textit{MEC Server Offloading Optimization:}
After obtaining $f^{loc}_m, \forall m \in \mathcal M$, UD-RRB-AP scheduling, and  power control $P_n, \forall n \in \mathcal N$, problem $\mathcal P_1$ reduces to
\begin{subequations}
	\begin{align} \nonumber 
	& \mathcal{P}_6: 
	\min_{\substack{\mathbf X, \bf y}} \pi\\
	&\rm s.t. \quad \text{C1, C4, C8.}
	\end{align}
\end{subequations}
More precisely, this optimization considers admission control and offloading decisions for MEC servers. In the next two sections, we present  efficient methods to solve the optimization problems $\mathcal{P}_2$ and $\mathcal{P}_6$ using conflict graph models.

\section{A Graph Theory-based Solution: Joint Approach}\label{G}
In this section, we develop an efficient joint solution to the 
optimization problem $\mathcal P_2$ and $\mathcal P_6$ using techniques inherited from graph theory. To this end, we
first design a joint MEC graph, denoted by J-MEC graph, that represents all feasible UD and power allocation
schedules. As such, $\mathcal P_2$ can be solved jointly. Given the solution of $\mathcal P_2$, we then develop admission control and offloading decision algorithm for solving $\mathcal P_6$.

\subsection{J-MEC Graph Design and Power Optimization}\label{J-MEC}
\textit{1) J-MEC graph description:} Let $\mathcal{A}$ denote the set of all possible combinations between UDs, APs, and RRBs, i.e., $\mathcal{A}=\mathcal{U}\times \mathcal{Z} \times\mathcal K$, and $a$ is a NOMA association which is an element in  $\mathcal{A}$, i.e., $a\in \mathcal A=\{n^a_1, n^a_2, z^a, m^a\}$. For convenience, $n^a$ represents the $n$-th UD in  association $a$. 
The weighted undirected J-MEC graph is denoted by $\mathcal{G}_\text{J-MEC}(\mathcal V, \mathcal E, \mathcal W)$ where $\mathcal V$ stands for the set of all the vertices, $\mathcal E$ is the set of all the edges, and $\mathcal W$ denotes the set of vertex weights. The designed J-MEC graph graph considers all the conflict transmissions between UDs across all RRBs in all APs. A vertex $v =\{n_1^v, n_2^v, z^v, m^v\}\in \mathcal{V}$ in this graph is generated for each association in $\mathcal A$, i.e., $v=a$ and $|\mathcal V|=|\mathcal A|$. Two distinct vertices $v_{i}$ representing $a$ and $v_{j}$ representing $a'$ are adjacent by a scheduling conflict edge if one of the
following cases occurs:
\begin{itemize}
	\item \textbf{CC1:} The same	UDs (any UD or both UDs) are associated with both vertices $v_{i}$ and $v_{j}$.
	\item \textbf{CC2:} The same RRB in the same AP (or different APs) is associated with both vertices $v_{i}$ and $v_{j}$.
\end{itemize} 
Mathematically, two distinct vertices $v_{i}$ representing $a$ and $v_{j}$ representing $a'$ are connecting by a conflict edge if and only if $a\cap a' \neq \emptyset$.
	
To select the UD-RRB-AP scheduling that provides a local minimum delay and guarantees minimum energy consumption, we assign a weight
$w(v)$ to each vertex $v \in \mathcal{G}_\text{J-MEC}$.  For notation simplicity, we define the utility
of UD $n$ as $X_n=U^n_{z,m}+\frac{B_n\lambda_n}{f^\text{loc}_m}+E^\text{loc}_{t_n}$. Therefore, the weight of vertex $v$ that reflects
both the minimum delay and energy consumption can be given by 
\begin{equation}\label{W(v)}
w(v)= X_{n_1^v}(p^*_{n_1^v}, p^*_{n_2^v}, z^v, m^v)+X_{n_2^v}(p^*_{n_1^v}, p^*_{n_2^v}, z^v, m^v)
\end{equation}
where $X_{n_1^v}$ and $X_{n_2^v}$ are the utility of two UDs $n_1^v$ and $n_2^v$, respectively. The weight of vertex $v$ in \eref{W(v)} is determined by the transmit powers $p^*_{n_1^v}$, $p^*_{n_2^v}$, RRB $z^v$, and AP $m^v$ allocated to them.

Using $\mathcal{G}_\text{J-MEC}$,  the optimization problem $\mathcal P_2$ for a fixed $\bf f^\text{loc}$ is similar to
minimum-weight independent set (MWIS) problems in several aspects. In  MWIS problems, two vertices must be nonadjacent in the graph, and similarly, in problem $\mathcal P_2$, two NOMA clusters cannot be allocated with the same RRB or contain
at least one UD. Moreover, the objective of problem $\mathcal P_2$ is to minimize the delay and energy consumption, and similarly, the
goal of MWIS is to minimize the weight of all vertices. Consequently, we have the following theorem.\\ 
\textit{\textbf{Theorem}}
\textit{Using $\mathcal{G}_\text{J-MEC}$, problem $\mathcal P_2$ for fixed $\bf f^\text{loc}$ can be equivalently transformed to the problem of determining the MWIS.}

\begin{proof} 
Let $\Gamma^*=\{v_{1},v_{2}, \cdots, \, v_{|\Gamma|}\},~ \forall v \in \mathcal{G}_\text{J-MEC}$, be the MWIS that  is associated with the feasible schedule $\{ \{a_1,a_2,\cdots, \, a_{|Z|}\},\ \cdots, \, \{a'_1,a'_2,\cdots, \, a'_{|Z|}\} \}$. Let $\Gamma$ is the set of all possible independent sets in $\mathcal{G}_\text{J-MEC}$. For each vertex $v \in \Gamma^*$ that is associated with association of UD, power, rate, RRB, and AP, the weight $w(v)$ in \eref{W(v)} is the minimum local computation delay and energy consumption that the induced NOMA cluster in vertex $v$ receives, i.e., the utility under the optimal transmit power of UDs. Therefore, the weight of the MWIS $\Gamma^*$  is precisely the objective function of problem $\mathcal P_3$ and can be written as $ w(\Gamma^*)= \sum\limits _{v\in \Gamma^*}w(v)= \sum\limits _{a\in \mathcal{A}}w(a)$.  Since each vertex is a feasible NOMA cluster, i.e., same UDs are scheduled to different RRBs, constraints (C2) and (C3) hold. 
\end{proof}

\ignore{Essentially, any maximal IS in $\mathcal{G}_\text{J-MEC}$ represents UD scheduling and power control that satisfies the following criterion.
\begin{itemize}
	\item The represented UDs by vertices in the MWIS are scheduled to the best RRBs/APs so as to transmit their tasks quickly. 
	\item The power level of UDs in each NOMA cluster is optimal. This results in good transmission rate from UDs to RRBs, and thus reduces the uplink transmission delay for transmitting the tasks to the APs.
\end{itemize}}

\textit{2) Power control optimization:}
A proper power allocation for each UD leads to  suppress the interference in NOMA clusters, thus a better uplink transmission rate is achieved. As a result, the uplink transmission duration for delivering tasks to RRBs/APs is minimized. Consider a NOMA-cluster in $\mathcal G_\text{J-MEC}$ that is associated with a feasible scheduling $a=\{n^a_1, n^a_2, z^a, m^a\}$. Our goal is to obtain a local optimal UD power allocation 
vector, denoted as  ($p^{*}_{n_1},p^{*}_{n_2}$) for that NOMA-cluster. The power allocation problem
is formulated as an optimization problem of maximizing the weighed sum-rate. As such, all the scheduled UDs transmit
their tasks to the associated RRBs/APs with minimum uplink transmission duration, which can be expressed as follows
\begin{subequations}
	\begin{align} \nonumber 
	&\mathcal{P}_7:  \max_{p_{n_1},p_{n_2}}\sum _{i=1}^2  \min_{n_i \in a} \log_{2}(1+\gamma^{n_i}_{m,z}),\\
	&\rm s.t. \quad
	0\leqslant P_{n_i}\leqslant P_{\max},~\forall~ n_i\in a, 
	\end{align}
\end{subequations}
where the optimization is over the power levels $p_{n_1},p_{n_2}$.

\subsection{MEC Tasks Offloading}
After obtaining the UD scheduling and $\bf f^\text{loc}$ of all APs in the network, by solving $\mathcal P_2$, now we are ready to solve $\mathcal P_6$ for tasks offloading. As mentioned before, the number of APs is higher than the number of MEC servers, thus there can be at most $K$ collected tasks admitted in and served by the MEC servers. Accordingly, we need to perform admission control, to pick out the number of collected tasks
$K$. Intuitively, APs in good channel conditions will benefit from task offloading, since the data transmission rate will be high, and consequently low delay in data offloading transmission. Moreover, the characteristics of tasks can play important roles. In the one hand, tasks with high sizes are most likely not suitable for offloading as the energy consumed in uplink transmission may be higher than the energy consumed in task processing. On the other hand, tasks with high processing density can be beneficial in task offloading to MEC servers as less energy will be consumed in task processing. 

Inspired by the aforementioned observations, we propose a \textit{cross-layer} weighting solution for admission control that are explained as follows. For each AP, we define the \textit{first-layer} weight as follows
\begin{equation} \label{Gm}
g^{loc}_m=T^\text{loc}_{\tau_{m}}+E^\text{loc}_{\tau_{m}}.
\end{equation}
In \eref{Gm},   $g^{loc}_m$ represents both the latency and consumption energy if collected tasks $m$ has been processed locally, given the solution to $\mathcal P_2$.  Now, let us define the \textit{second-layer} weight as follows
\begin{equation} \label{Gm2}
G^{mec}_{m,k}=\frac{C^m_k \lambda_m}{B_m},
\end{equation}
where $C^m_k$ indicates the transmission data rate from the $m$-th AP to the $k$-th MEC server. Particularly, a larger primary value of $G^{mec}_{m,k}$ offers a smaller uplink time from the AP of each represented task $m$. The larger $C^m_k$ is, the more likely it is for the $m$-th AP to perform task offloading. Based on this,
we propose low complexity algorithm that captures admission control among the local-infeasible AP. We first sort $g^{loc}_m$ in a decreasing order and select
the first $K$ APs to perform task offloading based on the maximum $G^{mec}_{m,k}$, and
the rest will do local processing.
\begin{algorithm}[t!]
	\begin{algorithmic}[1]
		\STATE \textbf{Require:} UD scheduling, power control $\bf P$, and local computations of all APs $\bf f^\text{loc}$.\;
		\FOR{$m\in \mathcal M$}
		\STATE Calculate $g^{loc}_m= T^\text{loc}_{\tau_{m}}+E^\text{loc}_{\tau_{m}}.$
		\FOR{$k\in \mathcal K$}
		\STATE Calculate $G^{mec}_{m,k}=\frac{C^m_k \lambda_m}{B_m}$.
		\ENDFOR
		\ENDFOR
		\STATE Sort all the APs in $\mathcal M$ in a decreasing order $\textbf{\textit{I}}(m)=\textit{I}(1),\textit{I}(2),\cdots, \textit{I}(M)$ based on $g^{loc}_m$.
		
		\FOR{$m=\{1,2, \cdots, M\}$}
		\IF{$m<K$}
		
		\STATE Find AP-MEC server scheduling based on maximum second-layer weight, i.e., $\max_{k\in \mathcal K}G^{mec}_{m,k}$.\; 
		\STATE Set $\mathcal K \leftarrow \mathcal K\backslash k$.\;
		\STATE Set $y_m=1$.\;
		\ELSE
		\STATE Set $y_m = 0$.
		\ENDIF
		\ENDFOR
		
		\STATE \textbf{Output:} $\bf y$.
	\end{algorithmic}
	\caption{Admission Control Algorithm} \label{alg1}
\end{algorithm}

\subsection{Greedy Algorithm}
The optimization problem $\mathcal P_2$ is a mixed-integer non-linear programming problem. The global solution is, therefore, equivalent to a minimum-wight independent set over J-MEC graph, which is NP-hard problem \cite{39}, and so is the problem $\mathcal P_2$. However, such problem can be near optimally solved with a reduced complexity as compared the $\mathcal{O}(|\mathcal{V}|^2.2^{|\mathcal{V}|}))$ naive exhaustive search methods, e.g., the algorithm in \cite{40}. The MWIS can be solved efficiently as explained in this subsection.  While the proposed solution is not necessarily optimum, it works very-well for solving  $\mathcal P_2$.

The joint local computation resource allocation and offloading decision optimization algorithm is broken into two phases as follows. 

\textit{\textbf{Phase I}}: In this phase, we solve the local computation resource allocation problem $\mathcal P_2$ in two stages (i) designing the J-MEC graph and the minimum weighted vertex search algorithm and (ii) finding the local computations of APs. This phase is explained as follows.

\textbf{Stage 1:} First, the J-MEC graph can be designed as follows. We generate all the possible schedules $\mathcal A$ of UD-NOMA clusters, RRBs, and APs. Afterwards, for each feasible schedule $a\in\mathcal{A}$, a vertex $v\in\mathcal{G}_\text{J-MEC}$ is generated. The optimal power levels of each association are  calculated by solving the optimization problem $\mathcal P_7$. The vertex in $v\in\mathcal{G}_\text{J-MEC}$ is created by appending the computed power levels and the corresponding rates to that vertex. We repeat the same steps above for all vertices. The J-MEC graph is, then, constructed by adding connections according to \textbf{CC1} and \textbf{CC2}.\\ 
Second, the algorithm itratively and greedily selects the MWIS $\Gamma^*$ among all the maximal independent sets $\Gamma$ in the J-MEC graph, where in each iteration we implement the following procedures. The algorithm computes the weight of all generated vertices using \eref{W(v)}. The vertex with the minimum weight $v^*$ is selected among all other corresponding vertices. The selected vertex $v^{*}$ is, then, added to $\Gamma^*$, where $\Gamma^*$ is initially empty. Afterwards, we update the $\mathcal{G}_\text{J-MEC}$ graph by removing the selected vertices $v^{*}$ and its connected vertices. As such, the next selected vertex is not in conflict connection with the already selected vertices in $\Gamma^*$. The process continues until no more vertices exist in J-MEC graph $\mathcal{G}_\text{J-MEC}$. Since each RRB in each AP contributes by a single vertex, the number of vertices in $\Gamma^*$ is $Z_\text{tot}$. 

\textbf{Stage 2:} Given the UD scheduling and power allocation of the APs from stage 1, we now find the local computation resources of the APs as follows. We first calculate the local computations of the collected tasks in each RRB for all APs. Then, similar to \cite{38}, we repetitively perform the following three closed-form procedures.
\begin{enumerate}
\item If $\frac{(\sum^{|\tau_m|}_{n=1}B_n)\lambda_n}{|\tau_m|T_{\max}^n} < f_{\max}^{loc}$, the local processing of tasks $\tau_m$ is feasible, and most likely these collected tasks will be processed at AP $m$. Thus, we set $f^{loc}_m=\frac{(\sum^{|\tau_m|}_{n=1}B_n)\lambda_n}{|\tau_m|T_{\max}^n}$ and $x_m=0$.
\item If $\frac{(\sum^{|\tau_m|}_{n=1}B_n)\lambda_n}{|\tau_m|T_{\max}^n} = f_{\max}^{loc}$, the local processing is feasible, and most likely $\tau_m$ will be processed at AP $m$. Thus, we set $f^{loc}_m=\frac{(\sum^{|\tau_m|}_{n=1}B_n)\lambda_n}{|\tau_m|T_{\max}^n}=f_{\max}^{loc}$ and $x_m=0$.
\item If $\frac{(\sum^{|\tau_m|}_{n=1}B_n)\lambda_n}{|\tau_m|T_{\max}^n} > f_{\max}^{loc}$, local processing is infeasible, and most likely $\tau_m$ will be offloaded. Thus, we set $x_m=1$. Such collected tasks will not be considered for the next iteration of performing stage 1.
\end{enumerate}

The above two-stages process is repeated until a maximum number of iterations is reached. 

\textit{\textbf{Phase II}}: In the second phase, we solve the admission control and offloading decision optimization problem $\mathcal P_6$. Particularly, this phase characterizes the solution of $\mathcal P_6$ by allocating the collected tasks of APs to the MEC-servers, such that the delay and energy consumption of tasks offloading is minimized. We first calculate $g^{loc}_m, \forall m$ and $G^{mec}_{m,k}$ ($\forall m, k$), and then we sort the collected tasks in a descending order according to $g^{loc}_m$. The index of the sorted collected tasks is $\textbf{\textit{I}}(m)=\textit{I}(1),\textit{I}(2),\cdots, \textit{I}(M)$. The collected tasks with $\textit{I}(1)$
has the higher priority to be associated with the best available
MEC server. Each iteration is implemented as follows. We label the AP that has the maximum value of $g^{loc}_1$ and find its corresponding second-layer weight that has the maximum value $G^{mec}_{1,k}$ among all other corresponding AP-MEC server associations. The selected AP-MEC server association is, then, added to $\mathbf{I}$, where $\mathbf{I}$ is initially empty. Afterwards, we update the list of values $g^{loc}_m$ and $G^{mec}_{m,k}$ by removing the selected AP and its associated $g^{loc}_1$, set $y_1=1$, and set $\mathcal K \leftarrow \mathcal K\backslash k$. The algorithm, then, locates the second maximum-value $g^{loc}_{2}$ and find its corresponding AP-MEC server association that has the highest secondary value  $G^{mec}_{2,k}$.  The process continues until no more available MEC servers in the network, and the remaining APs will perform local processing computations. The process of phase II is presented in Algorithm \ref{alg1}.

The overall two-phase algorithm to the problem $\mathcal P_2$ and problem $\mathcal P_6$ is summarized in Algorithm \ref{alg2}.

\begin{algorithm}[t!]
	\begin{algorithmic}[1]
		\STATE \textbf{Require:} $\mathcal{N}, \mathcal{K}, \mathcal{Z}$, $h_{m,k}$ and $~h^n_{m,z}$, $(n,m,z)\in\mathcal{N}\times\mathcal{M}\times\mathcal{Z}$.\;
		\STATE \textbf{Repeat:}\;
		\STATE Initialize $\Gamma^* = \emptyset$. 
		\STATE \textbf{Solve $\mathcal P_2$} for fixed $\textbf f^\text{loc}$.\;
	  \STATE Design $\mathcal G_{\text{J-MEC}}$ according to \sref{J-MEC}.\;
		\FOR{\text{each}  $v\in \mathcal G_{\text{J-MEC}}$}
		\STATE 		Solve $\mathcal P_7$ to compute the optimal power allocations $\textbf{P}=\{p^*_{n_1^v}, p^*_{n_2^v}\}$.\;
		\STATE Obtain $v= \{(r^*_{n_1^v},p^*_{n_1^v},z^v, m^v),(r^*_{n_2^v},p^*_{n_2^v},z^v, m^v) \}$ according to $\textbf{P}$.\; \STATE Calculate $w(v)$ using \eref{W(v)}.\;
		\ENDFOR
		\STATE $\mathcal G_{\text{J-MEC}}(\Gamma^* ) \leftarrow \mathcal G_{\text{J-MEC}}$.\;
		\WHILE{$\mathcal G_{\text{J-MEC}}(\Gamma^* ) \neq \emptyset$}
		\STATE $v^*=\arg\min_{v\in \mathcal G_\text{J-MEC}(\Gamma )} \{w (v)\}$.\; 
		\STATE Set $\Gamma^* \leftarrow \Gamma^* \cup v^*$ and 	set $\mathcal G_{\text{J-MEC}}(\Gamma^* ) \leftarrow \mathcal G_{\text{J-MEC}}(v^*)$.\;
		\ENDWHILE
		\STATE \textbf{Solve $\mathcal P_6$ for the resulting $\Gamma^*$}.\;
			\FOR{$a=\{1,2,\ \cdots, \,|\Gamma^*|\}$}
		\STATE	Calculate $\frac{(\sum^{|\tau_m|}_{n=1}B_n)\lambda_n}{|\tau_m|T_{\max}^n}, \forall n\in a$.\;
		\IF{$\frac{(\sum^{|\tau_m|}_{n=1}B_n)\lambda_n}{|\tau_m|T_{\max}^n}< f_{\max}^{loc}$}
		\STATE Set $f^{loc}_{m}=\frac{(\sum^{|\tau_m|}_{n=1}B_n)\lambda_n}{|\tau_m|T_{\max}^n}$.\;
		\ELSIF{$\frac{(\sum^{|\tau_m|}_{n=1}B_n)\lambda_n}{|\tau_m|T_{\max}^n}= f_{\max}^{loc}$}
		\STATE Set $f^{loc}_{m}= f_{\max}^{loc}$.\;
		\ELSIF{$\frac{(\sum^{|\tau_m|}_{n=1}B_n)\lambda_n}{|\tau_m|T_{\max}^n}> f_{\max}^{loc}$}
		\STATE Label  these tasks for possible offloading process.\;
		
	   \ENDIF
		
	    	\ENDFOR
	   \STATE $t=t+1$.\;
	   \STATE \textbf{Stop} until $l=L_{max}.$
		\STATE Obtain $\Gamma^*$ and $\textbf f^\text{loc}$.
		\STATE \textbf{Solve} $\mathcal P_3$ for the resulting $\Gamma^*$ and $\textbf f^\text{loc}$ by execute Algorithm \ref{alg1}.\;
	\end{algorithmic}
	\caption{Joint Local Computation, UD Scheduling, and Power Optimization Algorithm} \label{alg2}
\end{algorithm}

\subsection{Complexity Analysis}
The computational complexity 
of Algorithm \ref{alg1} and Algorithm \ref{alg2} is analyzed as follows. 

\textit{1) Complexity of Algorithm \ref{alg1}:} The computational complexity of sorting all the APs in Step 8 of Algorithm \ref{alg1} is $\mathcal O\left(M\log_2M\right)$. Meanwhile, the required complexity of associating an AP with an MEC server is $\mathcal O\left(1\right)$, and the computational complexity for executing Steps 9-17 of Algorithm \ref{alg1} is $\mathcal O\left(K\right)$.  Hence, the total computational complexity of Algorithm \ref{alg1} is $\mathcal O\left(K+M\log_2M\right)$.

\textit{1) Complexity of Algorithm \ref{alg2}:} The computational complexity of generating all NOMA clusters representing all vertices in the J-MEC graph is $\mathcal O$$N\choose {2}$. Then, connecting all these vertices requires a complexity of  $\mathcal O$$N\choose {2}$$^2$. Therefore, the overall computational complexity is $\mathcal O$$N\choose {2}$$^2$. Such high complexity is due to generating all the possible NOMA clusters which increases significantly as the number of UDs in the network increases.

\section{A Graph Theory-based Solution: Pruning Graph Approach}\label{MT}
In the previous section, we solved $\mathcal P_2$ jointly for UD scheduling, power control $\bf P$, and local computation $\textbf f^\text{loc}$. This requires high computational complexity for building J-MEC graph and solving the power control optimization for each vertex. To tackle such high complexity, we recommend to solve $\mathcal P_2$ for fixed $\textbf f^\text{loc}$. In particular, our proposed innovative method in this section introduces a sequential pruning graph algorithm that judiciously generates NOMA clusters whose tasks are certainly can be processed locally at the APs while simultaneously designing a reduced J-MEC graph.  In the reduced J-MEC graph, we do not need to generate all the possible NOMA clusters in the network which significantly reduces its size.

Towards that goal, this section first addresses the optimization problem $\mathcal P_1$ as a UD scheduling, power control, and offloading decision optimization problem, and can be written as 
\begin{subequations}
	\begin{align} \nonumber 
	& \mathcal{P}_8: 
	\min_{\substack{\mathbf X, \mathbf C, \mathbf R, \bf y}} \pi\\
	&\rm s.t. \quad \text{C1, C2, C3, C4,  C6, C7, C8.}
	\end{align}
\end{subequations}
To solve the problem in $\mathcal{P}_8$, we develop a simple approach that first solves the UD scheduling and power optimization problem using the pruning graph method and then solves the admission control and offloading decisions as in Algorithm \ref{alg1}. 

\subsection{Low Complexity Graph Pruning Solution}
In this subsection, we propose a low complexity, yet suboptimal, solution for solving the
UD scheduling and power control problem part in $\mathcal P_8$. Particularly, we first check the condition for generating feasible vertices that their associated tasks can be processed locally. Based on this, we propose a method for generating only such  NOMA clusters while simultaneously constructing the reduced J-MEC graph.

\textit{1) Graph description:} Let $\mathcal G_\text{r}=(\mathcal V, \mathcal E, \mathcal W)$ represents the \textit{reduced} J-MEC graph. To design $\mathcal G_\text{r}$, we itertaively generate a vertex $v$ for each UD (UDs), RRB, and AP in the network as follows. We start from RRB $z=1$, and assume that UD $n=1$ is allocated to it. Then we calculate the local task processing computation $\frac{B_1\lambda_1}{T_{\max}^1}$ of $n=1$ and check the possible three scenarios:
\begin{enumerate}
	\item  If UD $n=1$ is infeasible for a local processing at the $z$-th RRB in the $m$-th AP, we suppose UD $n=2$ is associated with RRB $z$, and then continue to calculate $p^*_n$ and judge the	feasibility. 
	
	\item  If UD $n=1$ is feasible for a local processing at the $z$-th RRB in the $m$-th AP and $\frac{B_1\lambda_1}{T_{\max}^1}< \frac{f^{loc}_m}{Z}$, then we find the second UD $j=n+1$ (currently, $j = 2$), for the ($n = 1, z = 1$) pair. Afterwords, calculate the transmitting powers $p^*_n$ and $p^*_j$ and generate a vertex $v=\{(p^*_n,r^*_n, z,m), (p^*_j,r^*_j, z,m)\}$ that represents a NOMA cluster. We then compute the weight of that vertex $w(v)= X_{n}(p^*_{n}, r^*_{n}, z, m)+X_{j}(p^*_{j}, r^*_{j}, z, m)$ and update the graph $\mathcal G_\text{r}$. If adding $j=2$ is infeasible, we let
	$j=j+1=3$, and  we verify the 	feasibility and repeat the aforementioned step.  
	
	\item  If UD $n=1$ is feasible for a local processing at the $z$-th RRB and $\frac{B_n\lambda_1}{T_{\max}^1}= \frac{f^{loc}_m}{Z}$, then we allocate this UD to RRB $z=1$, calculate the transmitting power $p^*_n$, and generate a vertex $v=\{(p^*_n,r^*_n, z,m)\}$ that represents only one UD. We then compute the weight of that vertex $w(v)= X_{n}(p^*_{n}, r^*_{n}, z, m)$ and update the graph $\mathcal G_\text{r}$.
\end{enumerate}
By iteratively repeating the above process (1)-(3) for
all $j\in \mathcal N$, $j > n$, we can obtain all the feasible vertices $(n=1, j\in \mathcal N, z=1), j > n$. To obtain all the feasible NOMA clusters, we repeat the above process for each $z, z\in \mathcal Z, m\in \mathcal M$. The vertices in the resulting constructed $\mathcal G_\text{r}$ are connected using \textbf{CC1} and \textbf{CC2} in section V.

\textit{2) Updated MWIS search method:} Since our proposed solution here greedily selects a number of UDs that can transmit their tasks to the RRBs/APs while minimizing the delay and energy consumption, we need to maximize the number of vertices that have minimum weights. In order to do that, the weight of each vertex needs to be updated. An appropriate design of the updated weights of vertices leads to selection of a large number of vertices and each vertex has minimum original weight that is defined in \eref{W(v)}. Such updated MWIS method was adopted in \cite{S1} and \cite{S3} to efficiently offload cloud and minimize delay, respectively. 

Let $\mathcal E_{v,v'}$ define the non-adjacency indicator of vertices $v$ and $v'$ in  the $\mathcal G_\text{r}$ graph such that:
\begin{equation}
\mathcal E_{v,v'} =
\begin{cases}
1 & \text{if $v$ is not adjacent to $v'$ in $\mathcal G_\text{r}$}, \\
0 & \text{otherwise}.
\end{cases}
\end{equation}
Next, let $\Delta_v$ denotes the weighted degree  of vertex $v$, which can be defined by $\Delta_V = \sum_{v' \in \mathcal G_\text{r}} \mathcal E_{v,v'}. w(v')$, where $w(v')$ is the original weight  of vertex $v'$  defined in \eref{W(v)}.   Hence, the modified weight  of vertex $v$ is defined as
\begin{align}\label{eq15}
\psi (v) & = w(v) \Delta_v = w(v)\sum_{v'  \in \mathcal G_\text{r}} \mathcal E_{v,v'}. w(v').
\end{align}

In \eref{eq15}, the weight of a vertex $v$ has two
features: (i) it has a minimum original
weight and (ii) it is not connected to a large
number of vertices that have minimum original weights. Based on this, we iteratively and heuristically execute a greedy vertex search scheme as follows. Initially, we pick up a vertex $v^*$ that has the minimum weight $w(v^*)$ and add it to the maximal IS $\Gamma^*$ (i.e., $\Gamma^* = \{v^*\}$). Then,  the  subgraph $\mathcal G_\text{r}(\Gamma^*)$, which consists of vertices in graph  $\mathcal G_\text{r}$ that are not connected to vertex $V^*$, is extracted and considered for the next  selection. In the next step, a  new minimum weight  vertex $v'^*$ is selected from subgraph $\mathcal G_\text{r}(\Gamma^*)$ (at this point $\Gamma^* = \{v^*, v'^*\}$). We repeat this process until no further vertex is not connected  to all the vertices in $\Gamma^*$. This approach is presented in Algorithm \ref{alg3}.

\begin{algorithm}[t!]
	\begin{algorithmic}[1]
		\STATE \textbf{Require:} $ \bf f, \mathcal{N}, \mathcal{K}, \mathcal{Z}$, $h_{m,k}$ and $~h^n_{m,z}$, $(n,m,z)\in\mathcal{N}\times\mathcal{M}\times\mathcal{Z}$\;
		\STATE \textbf{Repeat:}\;
		\STATE Initialize $\mathcal G_\text{r}=\emptyset$. 
		\FOR{$m=1:M$}
		
		 \FOR{$z=1:Z$}
		  \STATE Set $n=1$
		  \STATE Calculate $\frac{B_n\lambda_n}{T_{\max}^n}$
		   \IF{$\frac{B_n\lambda_n}{T_{\max}^n}< \frac{f^{loc}_m}{Z}$}
		   \STATE Set $j=n+1$
		   \WHILE{$j<N$}
		    \IF{$\frac{(\sum_{i\in\{n,j\}}B_i)\lambda_n}{2T_{\max}^i}\leq \frac{f^{loc}_m}{Z}$}
		    \STATE Calculate $p^*_n$ and $p^*_j$ according to $\mathcal P_7$.
		    \STATE Generate vertex $v=\{(p^*_n,r^*_n, z,m), (p^*_j,r^*_j, z,m)\}$.
		    \STATE Set $\mathcal G_\text{r}\longleftarrow \mathcal G_\text{r}\cup v$.
		    \ENDIF
		    \STATE $j=j+1$.
		    \ENDWHILE
		    
		    \ELSIF{$\frac{B_n\lambda_n}{T_{\max}^n}= \frac{f^{loc}_m}{Z}$}
		    \STATE Set $p^*_n=P_\text{max}$.
		    \STATE Generate vertex $v=\{(p^*_n,r^*_n, z,m)\}$ and set $\mathcal G_\text{r}\longleftarrow \mathcal G_\text{r}\cup v$.
		    \STATE $n=n+1$

		    \ENDIF 
		\ENDFOR
		\ENDFOR
		\STATE For each generated vertex $v$, finds its neighborhood $\mathcal N_{\mathcal G}(v)$ according to \textbf{CC1}, \textbf{CC2}, and \textbf{CC3}.
		\STATE Calculate the weight of each vertex $w(v)$ as in \eref{W(v)}.
		\STATE Let $\Gamma^* = \emptyset, l=0, \mathcal G_l=\mathcal G_\text{r}$.\\
		\textbf{MWIS Search Method}
		\WHILE{$\mathcal V(\mathcal G_l)\neq \emptyset$}
		\STATE $v^*=\arg\min_{{v\in \mathcal G_l}(\Gamma)} \{w (v)\}$ and set $\Gamma \leftarrow \Gamma \cup v^*$.
		
		\STATE Let $\mathcal V(\mathcal G_{l+1})=\mathcal V(\mathcal G_l(\Gamma))$.
		\STATE $l=l+1$
		\ENDWHILE
		\STATE Output: The MWIS and get the corresponding $\mathbf S$ and $\bf P$.
	\end{algorithmic}
	\caption{Low Complexity Graph Pruning Algorithm} \label{alg3}
\end{algorithm}

\textit{\textbf{Remark 4:} Notably, the $\mathcal G_r$ graph contains only feasible clusters, and thus it is a sub-graph	of the J-MEC graph constructed in section V. Therefore, the designed $\mathcal G_r$ graph generated by Algorithm \ref{alg3} provides the near-optimal solution to $\mathcal P_2$.}

\vspace*{-0.3cm}
\subsection{Complexity Analysis}\label{CA}
The computational complexity of Algorithm \ref{alg3} is dominated by the required complexity of generating feasible NOMA clusters (i.e., vertices in the reduced J-MEC graph), and connecting the generated vertices. To generate the feasible NOMA clusters by executing Steps 4-24 of Algorithm \ref{alg3}, the required computational complexity is $\mathcal O\left(MZN\right)$. Meanwhile, the required complexity of connecting the generated vertices by executing Step 25 of Algorithm \ref{alg3} is $\mathcal O\left((MZN)^2\right)$.  Therefore, the overall computational complexity of Algorithm  \ref{alg3} is $\mathcal O\left(MZN+(MZN)^2\right) \approx \mathcal O\left(M^2Z^2N^2\right)$. Essentially, for a dense network with large number of UDs,  Algorithm \ref{alg3} requires significantly reduced computational complexity than the joint approach of Algorithm \ref{alg2}.

\section{Numerical Results}\label{NR}

\subsection{Simulation Setting and Comparison Schemes}
We consider a NOMA-enabled and multi-hop MEC system where APs and MEC servers have fixed locations and UDs are distributed randomly within a hexagonal cell of radius  $1500$m. Unless otherwise stated, we set the numbers of APs $K$ and MEC servers to $10$, $4$, respectively.  In addition, each UD has one task to be processed locally at APs or at MEC servers. The channel model follows the standard path-loss model, which consists of three components: 1) path-loss of $128.1+37.6\log_{10}(\text{dis.[km]})$ for UD-RRB/AP transmissions and path-loss of $148 + 40 \log_{10}(\text{dis.[km]})$ for AP-MEC server transmissions; 2) log-normal shadowing with $4$ dB standard deviation; and 3) Rayleigh channel fading with zero-mean and unit variance. The noise power and the maximum’ F-AP and  user power are assumed to be $-174$ dBm/Hz and $P_\text{max}=Q_\text{max}=-42.60$ dBm/Hz, respectively. The link bandwidth is $10$ MHz. Other parameters are summarized in Table \ref{table_2}. To assess the performance of our proposed joint and pruning graph approaches, we simulate various scenarios with different number of UDs $N$, input data $B_n$, number of RRBs $Z$, and processing density $\lambda_n$\ignore{, max tolerable latency $T^{\max}_n$, and max local process capability $f^{loc}_{\max}$}.  For the sake of comparison, our proposed schemes are
compared with the following baseline schemes.
\begin{itemize}
	\item \textbf{Local:} In this scheme, all APs process the  collected tasks
	locally, and local resource allocation optimization is performed.
	When AP local processing is  not feasible, unsuccessful
	task processing happens.
	\item \textbf{All-offload:} In this scheme, the APs offload their collected tasks
	to the MEC servers, and no local processing at the APs.  When MEC server side processing is  not feasible,
	unsuccessful task processing happens.
	\item \textbf{Random-offload:} In this scheme, resource allocation and tasks offloading decisions are made randomly, and other optimization is performed. For resource allocation, we pick up a random MWIS in the J-MEC graph.
	
\end{itemize}
Also, we adopt three performance metrics as follows:
(i) the \textit{latency-energy consumption cost function} that represents the objective in $\mathcal P_1$ for the proposed joint scheme and $\mathcal P_8$ for the proposed pruning graph scheme, (ii) the \textit{effective system capacity} that represents the total number of UDs whose tasks are successfully processed, and (iii) the \textit{latency} that was shown in \eref{latency}.

\begin{table}[t!]
	\renewcommand{\arraystretch}{0.9}
	\caption{ Simulation Parameters}
	\label{table_2}
	\centering
	\begin{tabular}{p{5.3cm}| p{2.5cm}}
		\hline
		\hline
		
		\textbf{Parameter} & \textbf{Value}\\
		\hline
		Cell radius & $1500$ m \\
		\hline
		Circle radius of AP's service area $\mathtt R$  & $750$ m \\
		\hline
		Cluster radius $R_{th}$  & $0.05$ Mbits/s \\
		\hline
		Input data size, $B_n$ & $[0.4, 0.6]$ Kbit\\
		\hline
		Processing density, $\lambda_n$ & $100$ \\
		\hline
		MEC server capability, $f_{mec}$ & $3$ G cycles/s \\
		\hline
		Local capability constraint, $f^{loc}_{\max}$ & $0.05$ G cycles/s\\
		\hline
		Maximum tolerable latency, $T_{\max}^n$ & $10$ ms\\
		\hline
		CPU architecture based parameter, $\alpha$ & $10^{-27}$\\
		\hline
		
	\end{tabular}
\end{table}

\begin{figure}[t!]
	\centering
	\begin{minipage}{0.494\textwidth}
		\centering
		\includegraphics[width=0.85\textwidth]{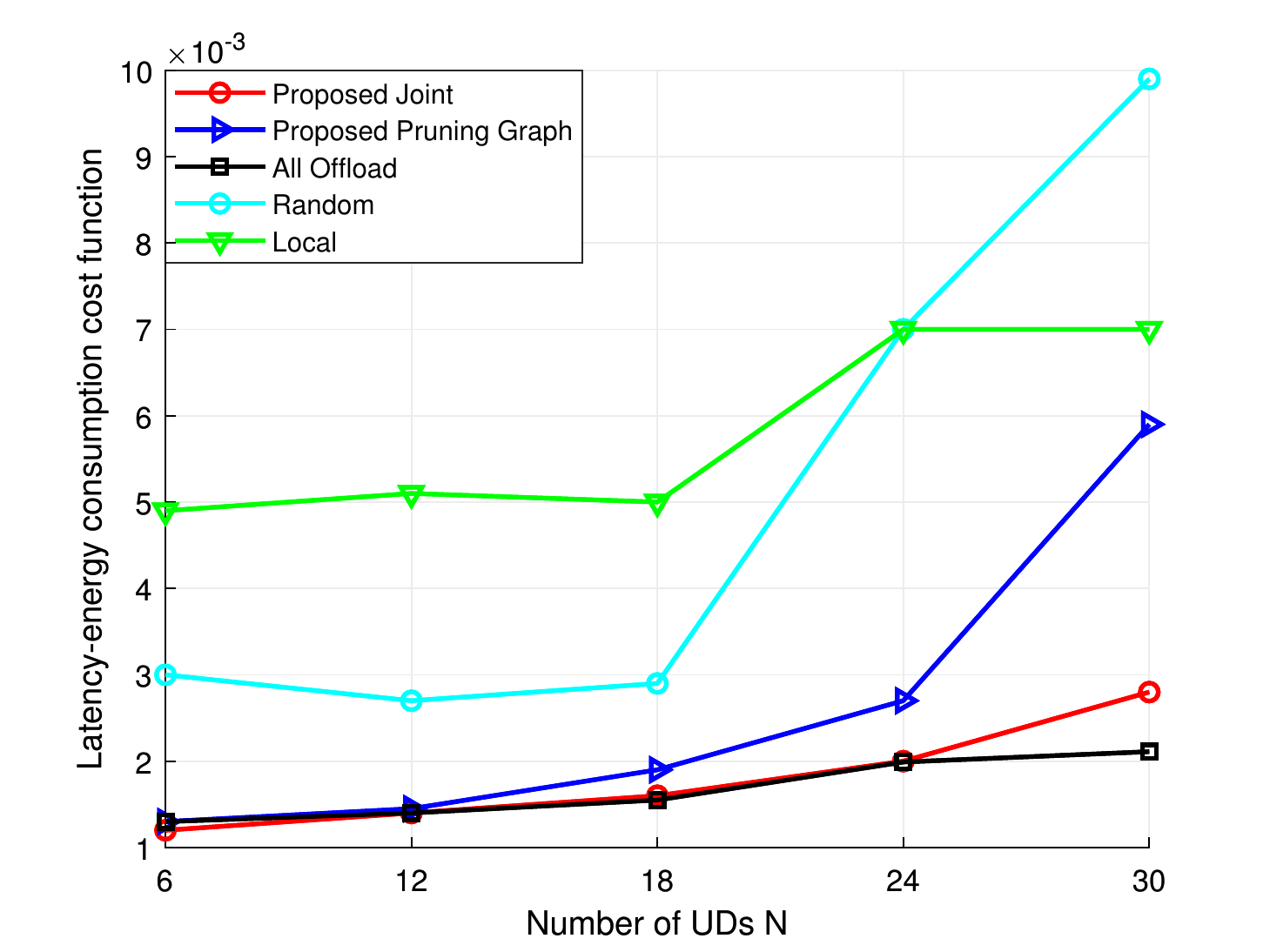} 
		\caption{Latency-energy consumption cost function vs. the number of UDs $N$ for $M=9$, $K=4$, and $Z=3$.}
		\label{fig5}
	\end{minipage}\hfill
	\begin{minipage}{0.494\textwidth}
		\centering
		\includegraphics[width=0.83\textwidth]{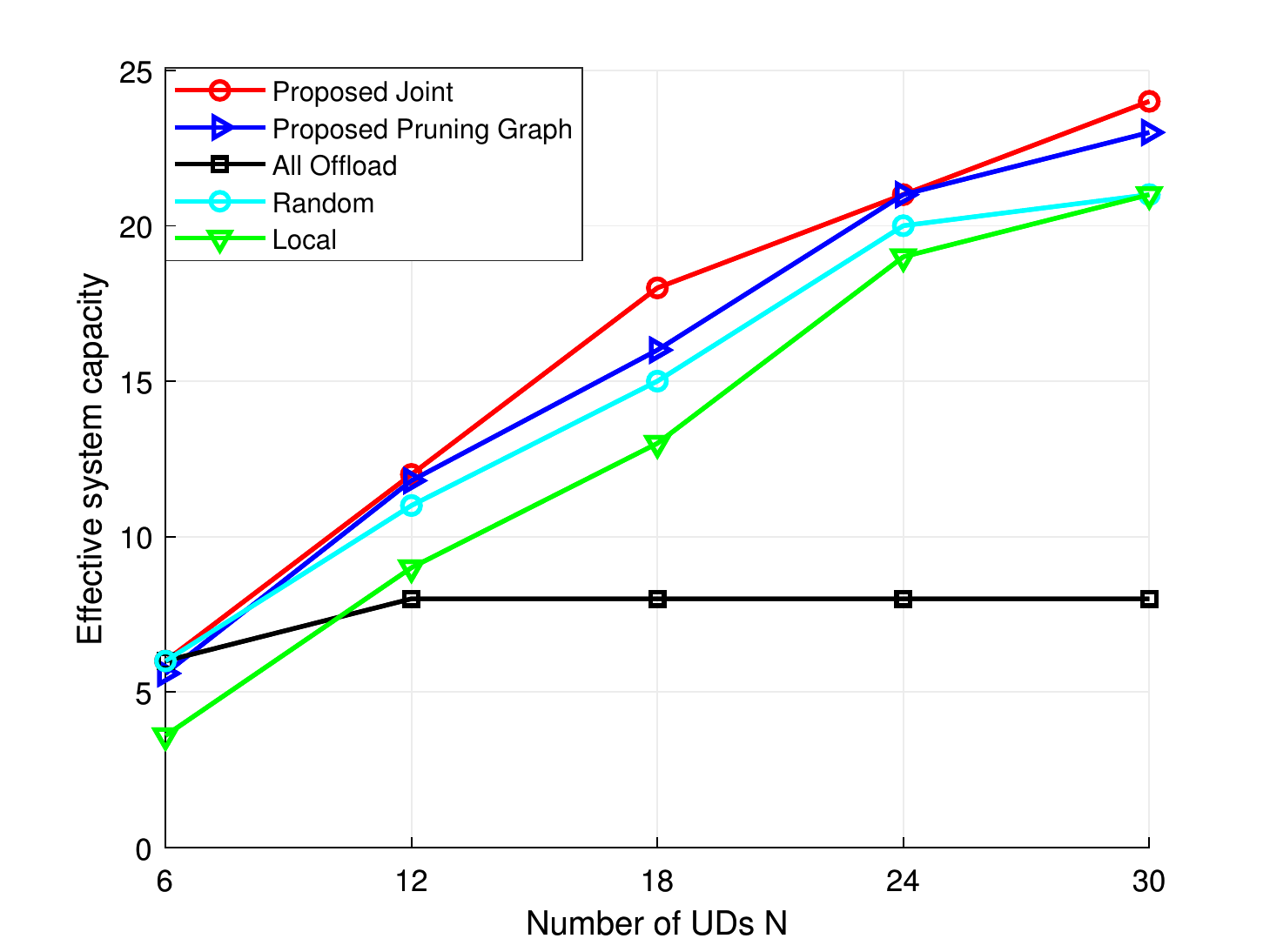} 
		\caption{Effective system capacity vs. the number of UDs $N$ for $M=9$, $K=4$, and $Z=3$.}
		\label{fig6}
	\end{minipage}\hfill
\end{figure}

We first plot in Fig. \ref{fig5} the latency-energy consumption cost function
versus the number of UDs $N$. From this figure, it can be seen that our  proposed schemes offer an improved performance in terms of cost function as compared to the other schemes. This improved performance is due to the joint 
and pruning graph schemes that (i) judiciously schedule UDs to APs/RRBs, adopt the transmission rate of each UD and optimize the transmission power of each UD, and (ii) smartly offload heavy intensive tasks that cannot be  locally processed at APs to the potential MEC servers.  Particularly, the random scheme suffers from randomly picking up a random MWIS that could have weak transmission rates from UDs and APs. As a result, a higher tasks uploading transmission, and it leads to a high latency. Further, the random selection of AP associations to MEC servers degrades its cost function performance. The local scheme focuses on processing the tasks locally at the APs, which degrades its cost function performance since APs have low processing capability. Thus, it consumes more energy and needs high latency for processing UDs' demanding tasks. On the other hand, in all-offload scheme where the collected tasks at the APs are offloaded, MEC servers have high processing capability, and accordingly, they can process the offloaded tasks quickly. This results in an improved performance as compared to all schemes, including our proposed pruning graph scheme. Notably, since all-offload scheme can only
benefit $N = 2K$ UDs, the cost function nearly
sightly changes when $N$ is greater than $8$. Our proposed joint scheme fully leverages the whole dimension of the J-MEC graph that considers a joint optimization of UDs scheduling, power control, and low processing optimization, and offloading decisions. Consequently, a close performance of our proposed joint scheme and all-offload scheme is achieved. This is because both local-related and MEC server-related optimizations
come into full play. Moreover, since the joint scheme considers all NOMA clusters, it works better than our proposed graph pruning scheme.

In Fig. \ref{fig6}, we plot the effective system capacity versus the
total number of UDs $N$. When $N$ increases form $6$
to $24$, the total number of RRBs across all APs is relatively sufficient ($Z=27$), so the system capacity grows relatively fast. The system capacity reachs $21$ supported UDs when $N = 24$
for the proposed schemes. When $N$ is nearly $30$, the
effective system capacity of our proposed schemes stop growing and can have at most $27$ supported UDs. Although the all-offload scheme has an improved cost function performance as in Fig. \ref{fig6}, it severely degrades the effective system capacity performance because it can serve at most $8$ UDs (i.e., $2K$). This makes the all-offload scheme impractical for dense NOMA-enabled and multi-hop MEC systems. The random scheme degrades the effective system capacity performance due to the random selection of MWIS in the reduced-NOMA graph, which results in a few number of vertices representing NOMA clusters. In contrast, our proposed schemes greedily select many vertices that have minimum weights and not adjacent to many vertices that have minimum weights. This shows the improved performance of our proposed schemes in Figs. \ref{fig5} and \ref{fig6}  as compared to the random scheme.

\begin{figure}[t!]
	\centering
	\begin{minipage}{0.494\textwidth}
		\centering
		\includegraphics[width=0.85\textwidth]{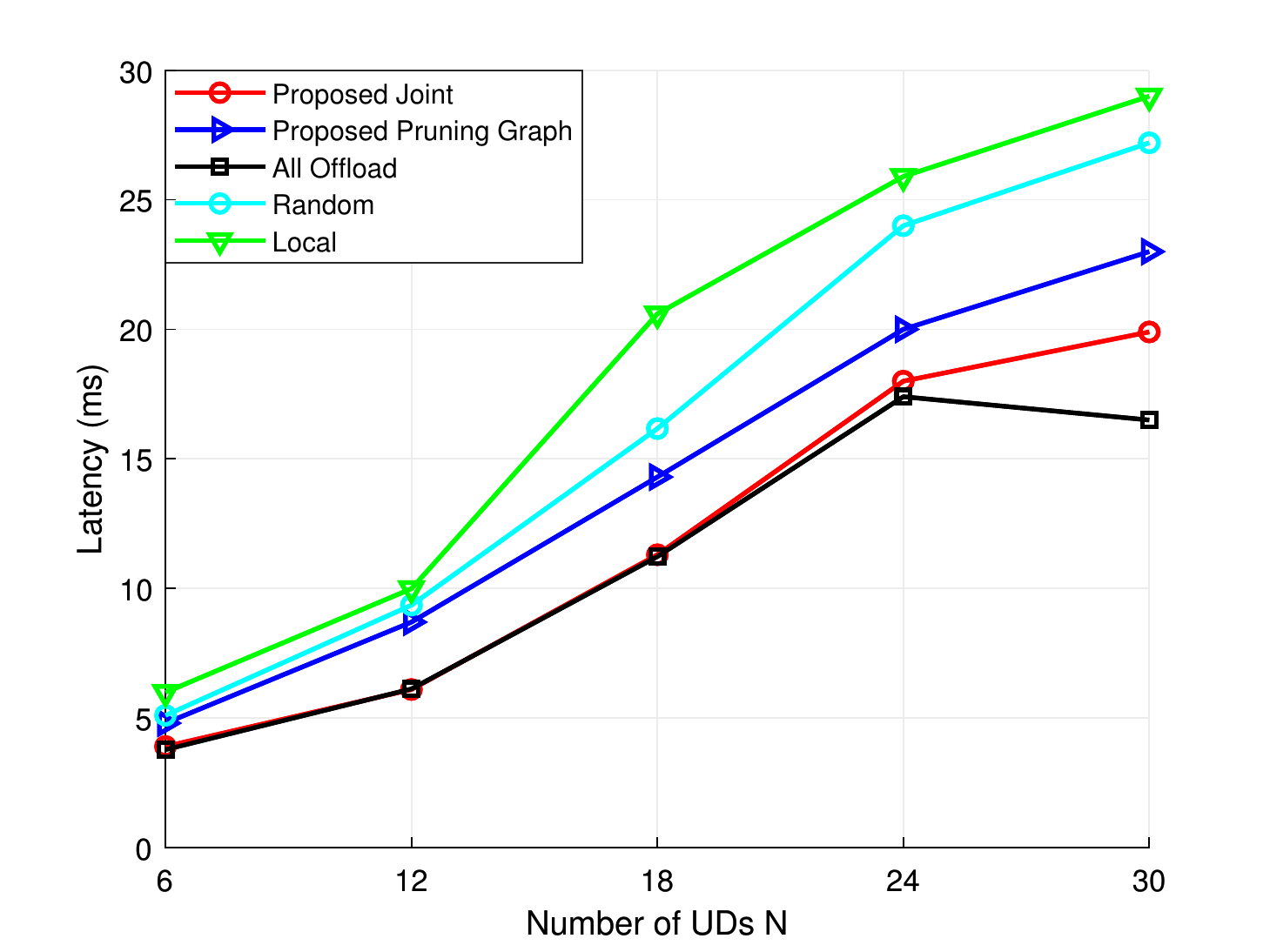} 
		\caption{Latency vs. the number of UDs $N$ for $M=9$, $K=4$, and $Z=3$.}
		\label{fig7}
	\end{minipage}\hfill
	\begin{minipage}{0.494\textwidth}
		\centering
		\includegraphics[width=0.85\textwidth]{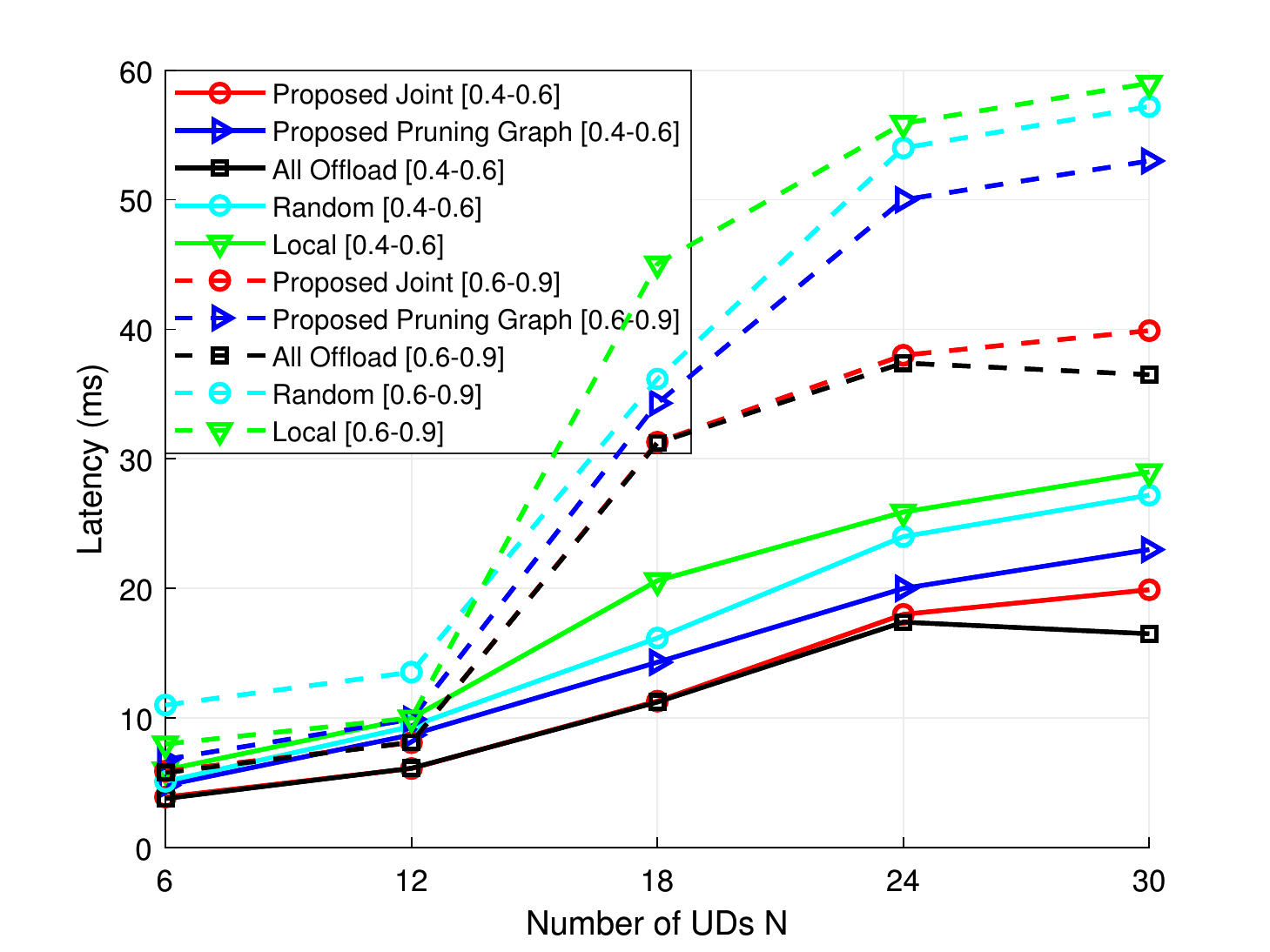} 
		\caption{Latency vs. the number of UDs $N$ for different input date $B_n$ for $M=9$, $K=4$, and $Z=3$.}
		\label{fig8}
	\end{minipage}\hfill
\end{figure}

In Fig. \ref{fig7}, we show the latency of processing UDs' tasks versus the number of UDs for an input data $B_n$ of $[0.4, 0.6]$ Kbit. Again, for
the above-mentioned reasons in Figs. \ref{fig5} and \ref{fig6}, our proposed schemes outperform other schemes. It can be observed from Fig. \ref{fig7} that increasing the number of UDs leads to an increased latency of all schemes. This is because when the number of UDs increases, the number of collected tasks for local processing or for offloading increases, thus leads to an increased in the maximum latency for uploading the tasks across all RRBs. To illustrate the impact of increasing the input data size $B_n$ on the latency, we plot in Fig. \ref{fig8} the latency against the number of UDs  for different ranges of $B_n$ of $[0.4, 0.6]$ and $[0.6, 0.9]$ Kbit. Fig. \ref{fig8} shows the size of input data and  how long it takes for the proposed solutions to upload and processed such data at APs and MEC servers. We can observe that the latency performances of all schemes increase with the data size. This is in accordance with the latency expression in \eref{latency}, where it was emphasized that  $\mathcal L(\mathbf X)$ increases with $B_n$. As $B_n$ increases, more bits are needed for uploading. Thus, time delay is increased to receive data from UDs. 

\begin{figure}[t!]
	\centering
	\begin{minipage}{0.494\textwidth}
		\centering
		\includegraphics[width=0.85\textwidth]{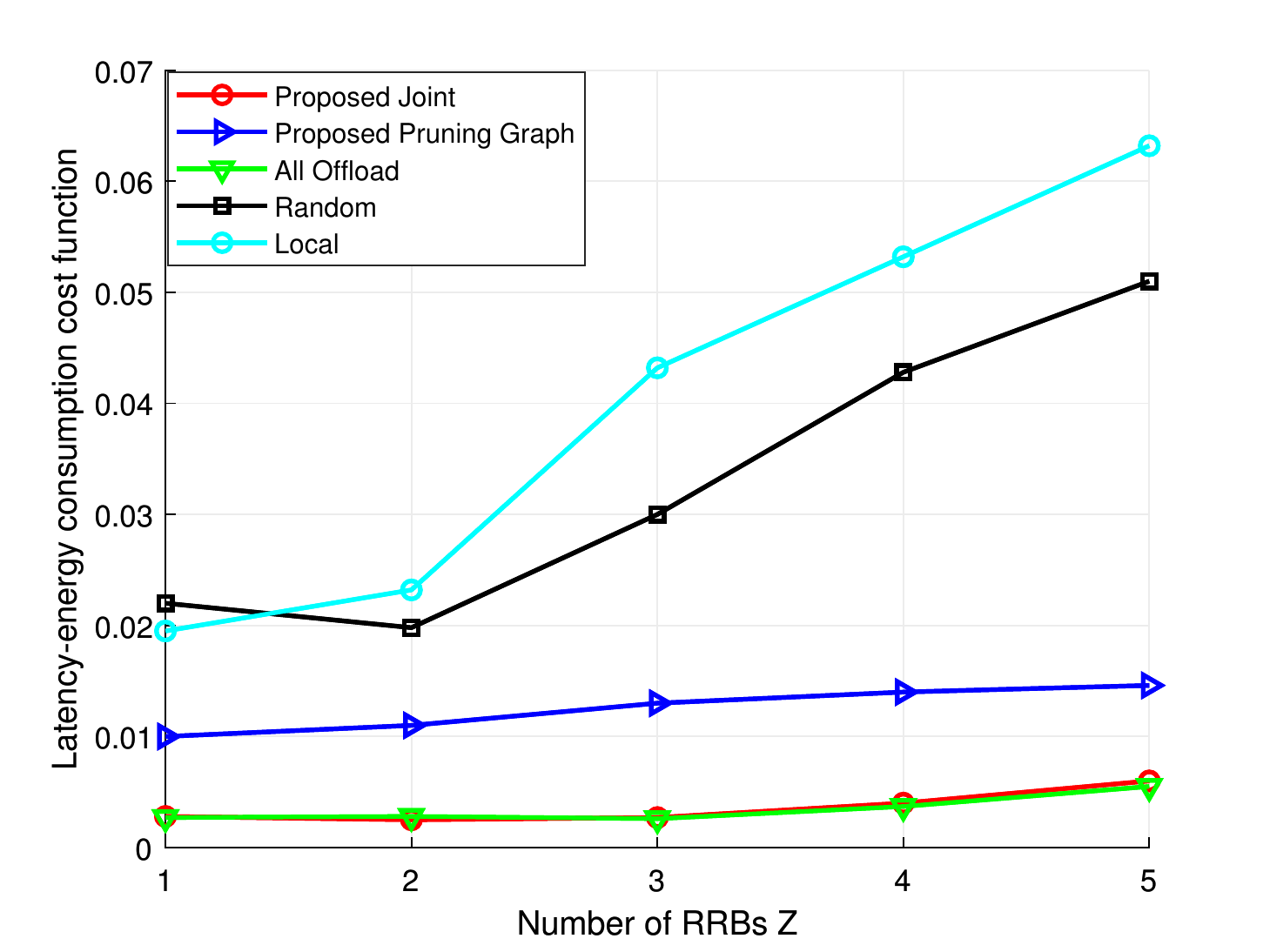} 
		\caption{Latency-energy consumption cost function vs. the number of RRBs $Z$ for $N=25$, $M=6$, and $K=3$.}
		\label{fig9}
	\end{minipage}\hfill
	\begin{minipage}{0.494\textwidth}
		\centering
		\includegraphics[width=0.85\textwidth]{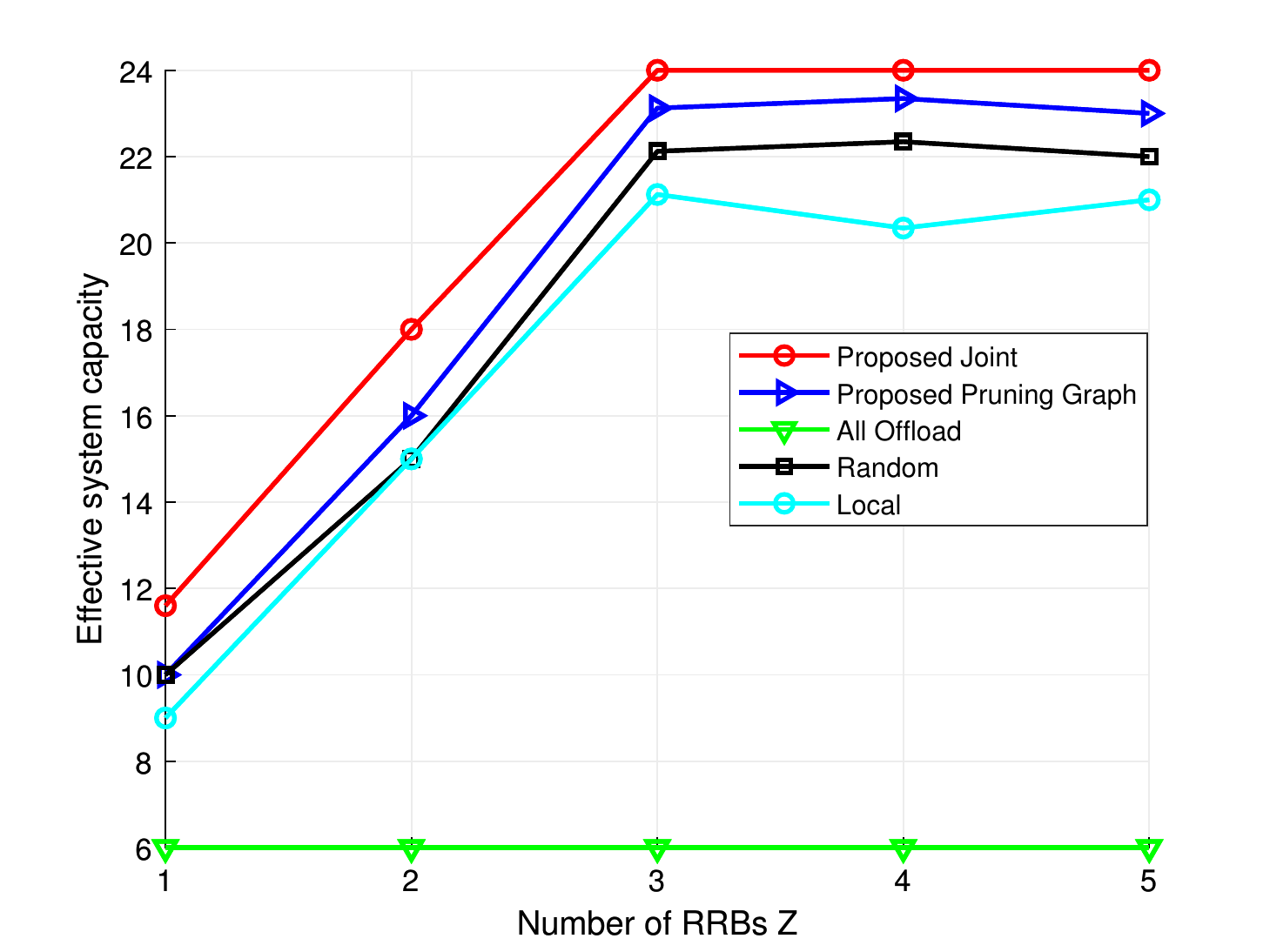} 
		\caption{Effective system capacity vs. the number of RRBs $Z$ for $N=25$, $M=6$, and $K=3$.}
		\label{fig10}
	\end{minipage}\hfill
\end{figure} 

In Figs. \ref{fig9} and \ref{fig10}, we plot the latency-energy consumption cost function and the effective system capacity versus the number of RRBs
$Z$, respectively. As can be seen, the number of UDs
that MEC servers and APs can afford increases linearly with $Z$. In all-offload
method, because UDs can offload their tasks
based on the number of MEC servers of $2K$, the cost function grows slowly and the effective
system capacity of the supported UDs keeps unchanged at $6$. Again, the all-offload scheme is not practical in terms of the effective system capacity, thus it serves in this work as a benchmark scheme. All other schemes, including our proposed, random, and local, follow the same rules, i.e., at first their effective system capacity grow fast, and then gradually slow  the number of RRBs. Meanwhile, it can also be found that the cost function and effective system capacity performances of our proposed algorithms always outperform the random and local methods.

\begin{figure}[t!]
	\centering
	\begin{minipage}{0.494\textwidth}
		\centering
		\includegraphics[width=0.85\textwidth]{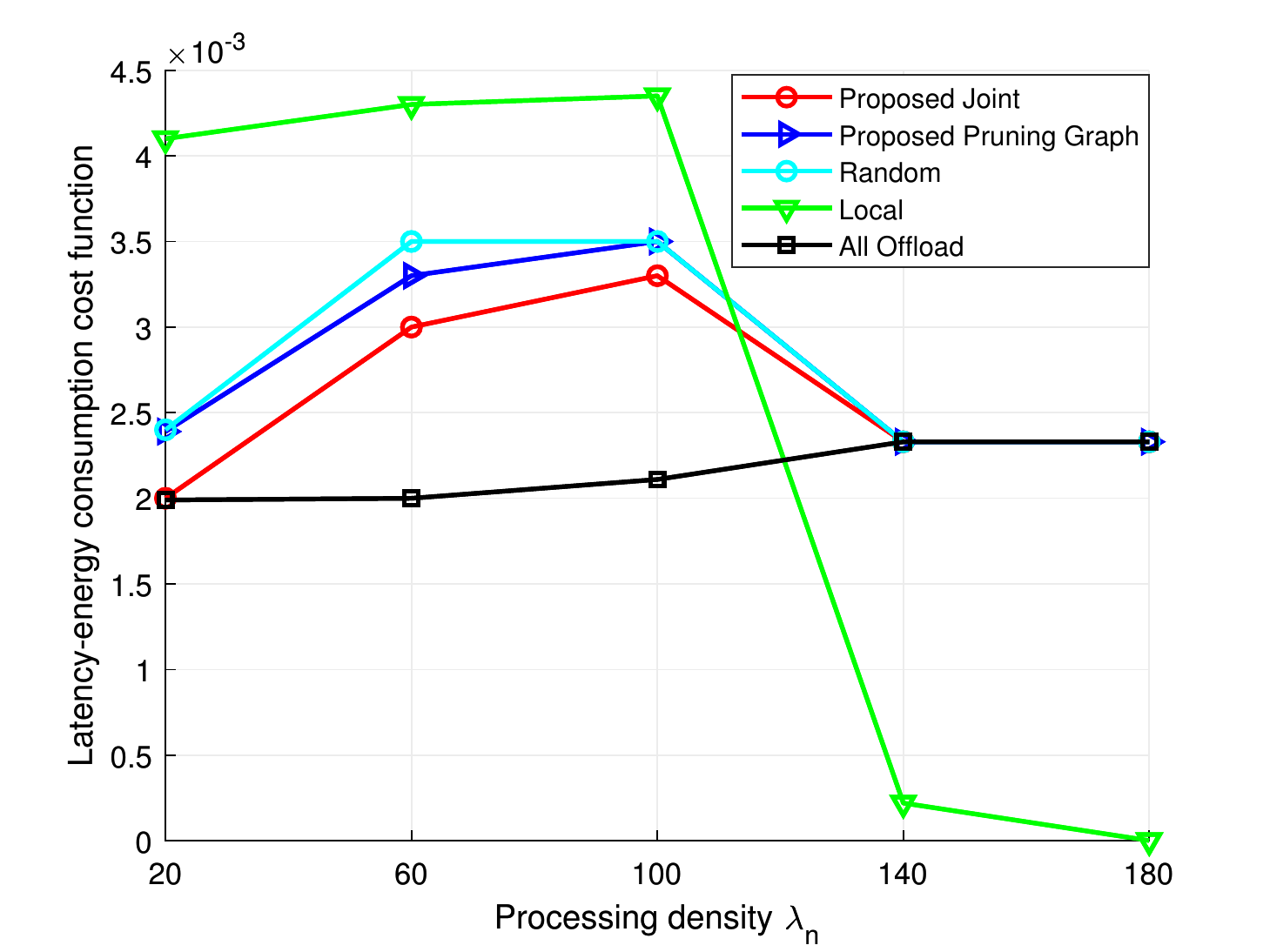} 
		\caption{Latency-energy consumption cost function vs. processing density $\lambda_n$ for $N=25$, $M=6$, and $K=3$, and $Z=2$.}
		\label{fig11}
	\end{minipage}\hfill
	\begin{minipage}{0.494\textwidth}
		\centering
		\includegraphics[width=0.85\textwidth]{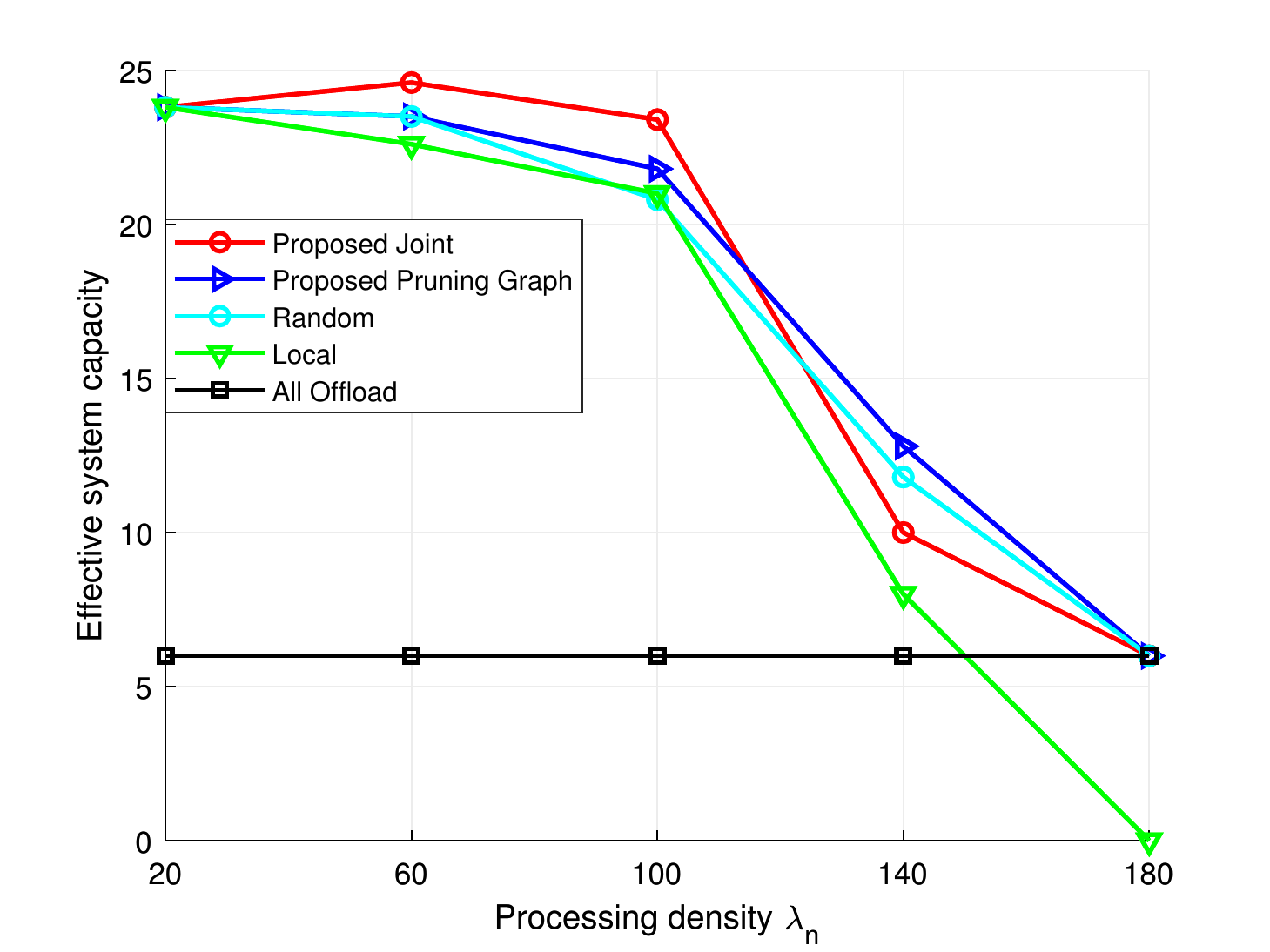} 
		\caption{Effective system capacity vs. processing density $\lambda_n$ for $N=25$, $M=6$, and $K=3$, and $Z=2$.}
		\label{fig12}
	\end{minipage}\hfill
\end{figure}

In Figs. \ref{fig11} and \ref{fig12}, we plot the latency-energy consumption cost function and the effective system capacity versus the processing density $\lambda_n$,
respectively, under default system size parameters, i.e., $N=25, M=6, K=3$, $Z=4$, and $B_n$ in the range of $[0.4, 0.6]$ Kbit. When $\lambda_n$ falls among a small numerical interval $[20, 60]$,
where the task is very simple and local processing is feasible
and suitable, almost all the $25$ tasks will be processed locally
with success. However, this slightly increases the cost function of the local scheme as the uploading transmission of such increased number of tasks is increasing.  Except for all-offload scheme, the
effective system capacity of other algorithms all reach the
maximum value of $25$ supported UDs and have the almost similar cost function performance.
For random scheme, since UDs and tasks are randomly
scheduled to UDs and offloaded to MEC servers, respectively, resulting in poor cost function performance and smaller effective system capacity, as shown in Figs. \ref{fig11} and \ref{fig12}. When $\lambda_n$ grows to $140$, the performance of local scheme
deteriorates rapidly, so the number of local feasible UDs
declines rapidly, i.e., when $\lambda_n=180$, the cost function of the local scheme is almost zero since no tasks of UDs can be locally processed at the APs. For our proposed schemes, due to multiple-dimensional joint optimization, the cost function slightly increases. In terms of effective system capacity,
since the number of local feasible UDs drops greatly, and
MEC servers can only accommodate $6$ UDs for task offloading, the effective capacity of all other algorithms reduce rapidly, except for all-offload scheme, as shown in Fig. \ref{fig12}.
Since the processing density $\lambda_n$ has a negligible effect on MEC server execution as in the all-offload scheme, the cost function slightly changes based on the latency of uploading tasks and the effective system capacity remains unchanged.

\section{Conclusion} \label{C}
In this paper, we investigated the joint optimization of latency and energy consumption in the NOMA-enabled and multi-hop MEC system  in which the APs are equipped with local processing  functionalities. By using the graph theory technique, we proposed two different approaches, namely, the J-MEC graph and the pruning graph approaches to obtain efficient solutions to the joint latency-energy optimization problem. The presented numerical results revealed that both proposed schemes achieve significant gains in terms of the latency and energy consumption minimization compared to the baseline solutions. Compared to the proposed J-MEC graph approach,  the pruning graph approach has some degradation in the system performance. However, this small performance degradation is obtained by reducing the computational complexity significantly compared to the joint approach. Therefore, our proposed  graph-based approaches offer a suitable trade-off between the performance and the computational complexity.

\begin{IEEEbiography}
	[{\includegraphics[width=1in,height=1.25in,clip,keepaspectratio, ]{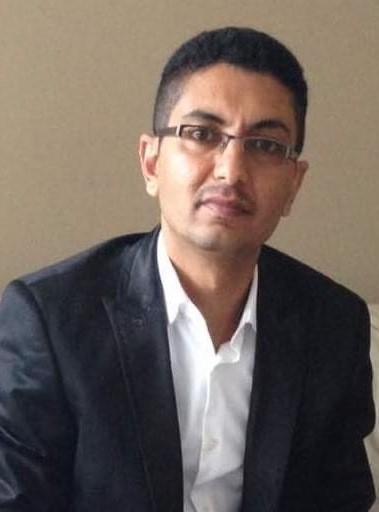}}]{Mohammed S. Al-Abiad} received the B.Sc. degree in computer and communications engineering from Taiz University, Taiz, Yemen, in 2010, the M.Sc. degree in electrical engineering from King Fahd University of Petroleum and Minerals, Dhahran, Saudi Arabia, in 2017, and the Ph.D. degree in electrical engineering from the University of British Columbia, Kelowna, BC, Canada, in 2020.  He is currently a Postdoctoral Research Fellow with the School of 	Engineering at the University of British Columbia, Canada. His research interests include  cross-layer network coding, optimization and
	resource allocation in wireless communication
	networks,  machine learning, and game theory. He is a student member
	of the IEEE.
\end{IEEEbiography}

\vspace{-1.5cm}	
\begin{IEEEbiography}
	[{\includegraphics[width=1in,height=1.25in,clip,keepaspectratio, ]{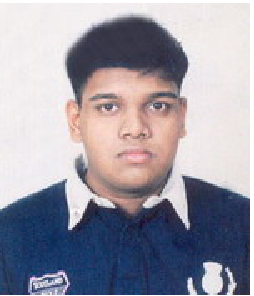}}]{Md. Zoheb Hassan}  received
	the Ph.D. degree from the University of British
	Columbia, Vancouver, BC, Canada, in 2019. He is
	a Research Fellows with the $\acute{\text{E}}$cole de technologie sup$\acute{\text{e}}$rieure (ETS), University of Quebec, Canada. His research interests include
	wireless optical communications, optimization and
	resource allocation in wireless communication
	networks, and digital communications over fading
	channels. He was the recipient of Four-Year Doctoral
	Fellowship of the University of British Columbia in
	2014. He serves/served as a Member of the Technical Program Committee of
	IEEE IWCMC 2018, IEEE ICC 2019, and IEEE ICC 2020.
\end{IEEEbiography}

\vspace{-1.5cm}	
\begin{IEEEbiography}
	[{\includegraphics[width=1in,height=1.25in,clip,keepaspectratio]{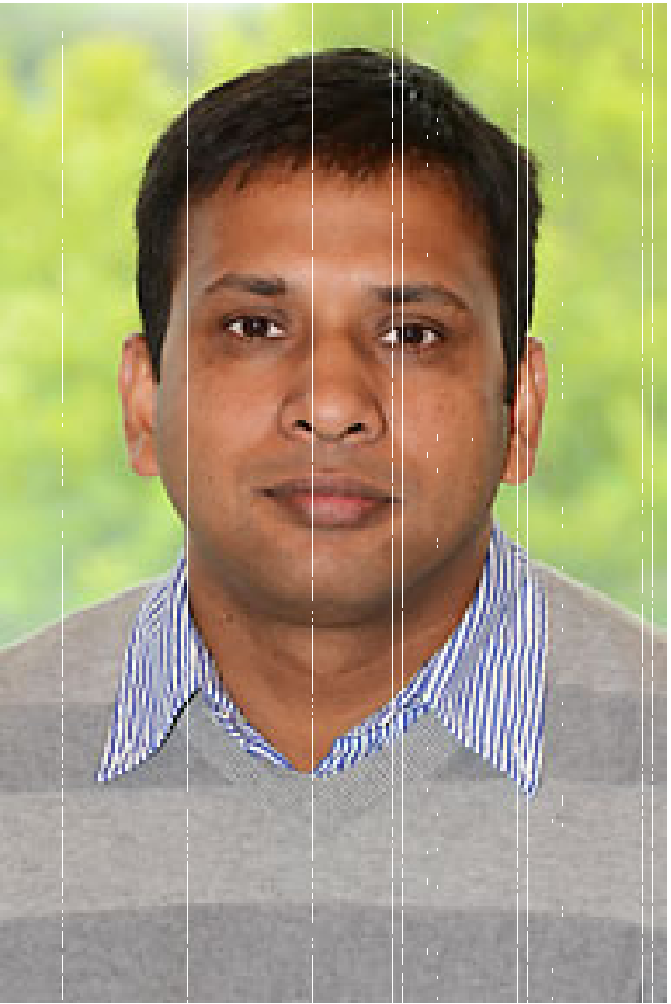}}]{Md. Jahangir Hossain}(S’04, M’08, SM’18) received the B.Sc. degree in electrical and	electronics engineering from the Bangladesh	University of Engineering and Technology (BUET),	Dhaka, Bangladesh, the M.A.Sc. degree from	the University of Victoria, Victoria, BC, Canada,
	and the Ph.D. degree from The University of
	British Columbia (UBC), Vancouver, BC. He was
	a Lecturer with BUET. He was a Research Fellow
	with McGill University, Montreal, QC, Canada,
	the National Institute of Scientific Research,
	Quebec, QC, and the Institute for Telecommunications Research, University
	of South Australia, Mawson Lakes, Australia. His industrial experience 	includes a Senior Systems Engineer position with Redline Communications,
	Markham, ON, Canada, and a Research Intern position with Communication 	Technology Lab, Intel, Inc., Hillsboro, OR, USA. He is currently an Associate Professor with the School of Engineering, UBC Okanagan campus, Kelowna,	BC. His research interests include designing spectrally and power-efficient	modulation schemes, applications of machine learning for communications,	quality-of-service issues and resource allocation in wireless networks, and
	optical wireless communications. He regularly serves as a member of the Technical Program Committee of the IEEE International Conference
	on Communications (ICC) and the IEEE Global Telecommunications	Conference (Globecom). He has been serving as an Associate Editor for
	IEEE COMMUNICATIONS SURVEYS AND TUTORIALS and an Editor for 	IEEE TRANSACTIONS ON COMMUNICATIONS. He previously served as an
	Editor for IEEE TRANSACTIONS ON WIRELESS COMMUNICATIONS.
\end{IEEEbiography}

\end{document}